\documentclass[useAMS,usenatbib,usegraphicx]{mn2e}

\usepackage{color}
\usepackage{multirow}

\renewcommand{\vec}[1]{\bmath{#1}}

\newcommand{\pd}[2]{\frac{\partial #1}{\partial #2}}
\newcommand{\DS}{\displaystyle}
\newcommand{\HALF}{\frac{1}{2}}
\newcommand{\gtrsim}{{\ga}}
\newcommand{\lesssim}{{\la}}

\title[HLLD solver for relativistic MHD]
      {A five-wave HLL Riemann solver for relativistic MHD}

\author[A. Mignone, M. Ugliano and G. Bodo]
       {A. Mignone$^{1,2}$\thanks{E-mail:
mignone@oato.inaf.it (AM)}, M. Ugliano$^{2}$ and G. Bodo$^{1}$\\
$^{1}$INAF/Osservatorio Astronomico di Torino, Strada Osservatorio 20, 10025 Pino Torinese, Italy\\
$^{2}$Dipartimento di Fisica Generale ``Amedeo Avogadro'' Universit\`a degli Studi di Torino, Via Pietro Giuria 1, 10125 Torino, Italy}

\begin{document}
\date{Accepted ??. Received ??; in original form ??}

\pagerange{\pageref{firstpage}--\pageref{lastpage}} \pubyear{2007}

\maketitle

\label{firstpage}

\begin{abstract}
 We present a five-wave Riemann solver for the equations of 
 ideal relativistic magnetohydrodynamics. Our solver can be regarded as
 a relativistic extension of the five-wave HLLD Riemann solver initially developed
 by Miyoshi and Kusano for the equations of ideal MHD.
 The solution to the Riemann problem is approximated by a 
 five wave pattern, comprised of two outermost fast shocks, two rotational
 discontinuities and a contact surface in the middle.
 The proposed scheme is considerably more elaborate than in the classical case 
 since the normal velocity is no longer constant across the rotational modes.
 Still, proper closure to the Rankine-Hugoniot jump conditions can be attained 
 by solving a nonlinear scalar equation in the total pressure variable
 which, for the chosen configuration, has to be constant over the whole Riemann fan.
 The accuracy of the new Riemann solver is validated against 
 one dimensional tests and multidimensional applications.
 It is shown that our new solver considerably improves over the popular
 HLL solver or the recently proposed HLLC schemes.
\end{abstract}


\section{Motivations}
%
%
%
%

Relativistic flows are involved in many of the high-energy
astrophysical phenomena, such as, for example, jets in extragalactic
radio sources, accretion flows around compact objects, pulsar winds
and $\gamma$ ray bursts. In many instances the presence of a magnetic
field is also an essential ingredient for explaining the physics of
these objects and interpreting their observational appearance. 

Theoretical understanding of relativistic phenomena is subdue to 
the solution of the relativistic magnetohydrodynamics (RMHD) equations
which, owing to their high degree of nonlinearity, 
can hardly be solved by analytical methods.
For this reason, the modeling of such phenomena has prompted
the search for efficient and accurate numerical formulations.
In this respect, Godunov-type schemes \citep{Toro97} have gained increasing
popularity due to their ability and robustness in accurately describing 
sharp flow discontinuities such as shocks or tangential waves.

One of the fundamental ingredient of such schemes is the exact or 
approximate solution to the Riemann problem, i.e., the decay between two 
constant states separated by a discontinuity.
Unfortunately the use of an exact Riemann solver
\citep{GR06} is prohibitive because of the huge
computational cost related to the high degree of 
nonlinearities present in the equations. 
Instead, approximate methods of solution are preferred. 

Linearized solvers \citep{K99, Balsara01, KKU02} rely on the 
rather convoluted eigenvector decomposition of the underlying 
equations and may be prone to numerical pathologies leading
to negative density or pressures inside the solution
\citep{EMRS91}.

Characteristic-free algorithms based on the Rusanov Lax-Friedrichs
or the Harten-Lax-van Leer \citep[HLL,][]{HLL83} formulations  
are sometime preferred due to their ease of implementation and 
positivity properties. Implementation of such algorithms can be
found in the codes described by \cite{GKT03, Leis05, dZZB07, vdHKM08}. 
Although simpler, the HLL scheme approximates only two out of the seven waves
by collapsing the full structure of the Riemann fan into a single 
average state. 
These solvers, therefore, are not able to resolve intermediate waves such as
Alfv\'en, contact and slow discontinuities. 

Attempts to restore the middle contact (or entropy) wave 
\citep[HLLC, initially devised for the Euler equations by][]{TSS94} 
have been proposed by \cite{MMB05} in the case
of purely transversal fields and by \cite{MB06} (MB from now on), \cite{HJ07} 
in the more general case.
These schemes provide a relativistic extension of the work
proposed by \cite{Gurski04} and \cite{Li05} for the classical
MHD equations.

HLLC solvers for the equations of MHD and RMHD, however, still do not 
capture slow discontinuities and Alfv\'en waves.
Besides, direct application of the HLLC solver of MB to genuinely 3D
problems was shown to suffer from a (potential) pathological singularity
when the component of magnetic field normal to a zone interface
approaches zero.

A step forward in resolving intermediate wave structures was then 
performed by \cite{MK05} (MK from now on) who, in the context of Newtonian
MHD, introduced a four state solver (HLLD) restoring the rotational (Alfv\'en) 
discontinuities. 
In this paper we propose a generalization of Miyoshi \& Kusano approach 
to the equations of relativistic MHD.
As we shall see, this task is greatly entangled by the different
nature of relativistic rotational waves across which the velocity 
component normal to the interface is no longer constant.
The proposed algorithm has been implemented in the PLUTO code 
for astrophysical fluid dynamics \citep{PLUTO} which embeds
a computational infrastructure for the solution of
different sets of equations (e.g., Euler, MHD or relativistic MHD 
conservation laws) in the finite volume formalism.

The paper is structured as follows: in \S\ref{sec:equations} we briefly 
review the equations of relativistic MHD (RMHD) and formulate
the problem. In \S\ref{sec:solver} the new Riemann solver is derived. 
Numerical tests and astrophysical applications are presented in 
\S\ref{sec:num} and conclusions are drawn in \S\ref{sec:conclusions}.

\section{Basic Equations}\label{sec:equations}
%
%
%

The equations of relativistic magnetohydrodynamics (RMHD) are derived 
under the physical assumptions of constant magnetic permeability and 
infinite conductivity, appropriate for a perfectly conducting fluid
\citep{Anile89, lich67}.
In divergence form, they express particle number and energy-momentum conservation:
\begin{eqnarray}
  \label{eq:clmass}
  \partial_\mu\left(\rho u^\mu\right) & = &  0 \,,
\\ \noalign{\medskip}
  \label{eq:clmomen}
  \partial_\mu\Big[\left(w_{\rm g} + b^2\right) u^\mu u^\nu 
   - b^\mu b^\nu 
  + \left(p_{\rm g} + \frac{b^2}{2}\right)\eta^{\mu\nu}\Big] & = & 0\,,
\\ \noalign{\medskip}
  \label{eq:clind}
  \partial_\mu\left(u^\mu b^\nu - u^\nu b^\mu\right) & = & 0\,,
\end{eqnarray}
where $\rho$ is the rest mass density, $u^\mu = \gamma(1,\vec{v})$ is
the four-velocity ($\gamma \equiv$ Lorentz factor, 
$\vec{v}\equiv$ three velocity), $w_{\rm g}$ and $p_{\rm g}$ 
are the gas enthalpy and thermal pressure, respectively.
The covariant magnetic field $b^\mu$ is orthogonal to the fluid four-velocity
($u^\mu b_\mu = 0$) and is related to the local rest frame
field $\vec{B}$ by
\begin{equation}\label{eq:bmu}
 b^\mu = \left[\gamma\vec{v}\cdot\vec{B}, 
               \frac{\vec{B}}{\gamma} + \gamma\left(\vec{v}\cdot\vec{B}\right)\vec{v}\right]\,.
\end{equation}
In Eq. (\ref{eq:clmomen}), 
$b^2 \equiv b^\mu b_\mu = \vec{B}^2/\gamma^2 + \left(\vec{v}\cdot\vec{B}\right)^2$ 
is the squared magnitude of the magnetic field.

The set of equations (\ref{eq:clmass})--(\ref{eq:clind}) 
must be complemented by an equation of state which
may be taken as the constant $\Gamma$-law: 
\begin{equation}\label{enth}
 w_g = \rho + \frac{\Gamma}{\Gamma-1}p_g\,, 
\end{equation} 
where $\Gamma$ is the specific heat ratio. Alternative equations
of state \citep[see, for example,][]{MmK07} may be adopted.


In the following we will be dealing with the one dimensional 
conservation law
\begin{equation}\label{eq:1dconslaw}
 \pd{\vec{U}}{t} + \pd{\vec{F}}{x} = 0\,,
\end{equation}
which follows directly from Eq. (\ref{eq:clmass})-(\ref{eq:clind}) 
by discarding contributions from $y$ and $z$.
Conserved variables and corresponding fluxes take the form:
\begin{equation}\label{eq:UandF}
  \vec{U} = \left(\begin{array}{c}
  D   \\ \noalign{\medskip}
  m^k \\ \noalign{\medskip}
  E   \\ \noalign{\medskip} 
  B^k \end{array}\right)
 \,,\quad
  \vec{F}  = \left(\begin{array}{c}
  Dv^x                            \\  \noalign{\medskip}
  wu^xu^k - b^xb^k + p\delta_{kx} \\ \noalign{\medskip}
  m^x \\ \noalign{\medskip}
  B^k v^x - B^xv^k\end{array}\right)
\end{equation}
where $k=x,y,z$, $D = \rho\gamma$ is the the density as seen from the
observer's frame while, introducing $w\equiv w_g + b^2$ (total enthalpy) 
and $p\equiv p_g + b^2/2$ (total pressure), 
\begin{equation}\label{eq:momE}
  m^k = wu^0u^k - b^0b^k \,,\quad
  E   = wu^0u^0 - b^0b^0 - p 
\end{equation}
are the momentum and energy densities, respectively.
$\delta_{kx}$ is the Kronecker delta symbol.

Note that, since $F_{B^x}=0$, the normal component of magnetic field
($B^x$) does not change during the evolution and can be regarded as a parameter.
This is a direct consequence of the $\nabla\cdot\vec{B} = 0$ condition.

A conservative discretization of Eq. (\ref{eq:1dconslaw}) 
over a time step $\Delta t$ yields
\begin{equation}\label{eq:1st_ord}
 \vec{U}_i^{n+1} = \vec{U}_i^n - \frac{\Delta t}{\Delta x}
                   \left(\vec{f}_{i+\HALF} - \vec{f}_{i-\HALF}\right)\,,
\end{equation}
where $\Delta x$ is the mesh spacing and $\vec{f}_{i+\HALF}$ 
is the upwind numerical flux computed at zone faces
$x_{i+\HALF}$ by solving, for $t^n < t < t^{n+1}$, the initial value 
problem defined by Eq. (\ref{eq:1dconslaw}) together with the initial condition
\begin{equation}\label{eq:riemann}
 \vec{U}(x,t^n) = \left\{\begin{array}{cc}
   \vec{U}_L & \quad\textrm{for}\quad x < x_{i+\HALF} \,, \\ \noalign{\medskip}
   \vec{U}_R & \quad\textrm{for}\quad x > x_{i+\HALF} \,, 
\end{array}\right. 
\end{equation}
where $\vec{U}_L$ and $\vec{U}_R$ are discontinuous left and right 
constant states on either side of the interface.
This is also known as the Riemann problem.
For a first order scheme, $\vec{U}_L = \vec{U}_i$ 
and $\vec{U}_R = \vec{U}_{i+1}$.

The decay of the initial discontinuity given by Eq. (\ref{eq:riemann})
leads to the formation of a self-similar wave pattern in the $x-t$ plane
where fast, slow, Alfv\`en and contact modes can develop.
At the double end of the Riemann fan, two fast magneto-sonic waves bound 
the emerging pattern enclosing two rotational (Alfv\`en) discontinuities,
two slow magneto-sonic waves and a contact surface in the middle.
The same patterns is also found in classical MHD.
Fast and slow magneto-sonic disturbances can be either shocks or rarefaction waves, 
depending on the pressure jump and the norm of the magnetic field.
All variables (i.e. density, velocity, magnetic
field and pressure) change discontinuously across a fast or a slow shock, 
whereas thermodynamic quantities such as thermal pressure and rest density 
remain continuous when crossing a relativistic Alfv\`en wave.
Contrary to its classical counterpart, however, the
tangential components of magnetic field trace ellipses instead of circles
and the normal component of the velocity is no longer continuous across 
a rotational discontinuity, \cite{K97}.
Finally, through the contact mode, only density exhibits a jump while 
thermal pressure, velocity and magnetic field remain continuous.

The complete analytical solution to the Riemann
problem in RMHD has been recently derived in closed form by \cite{GR06} 
and number of properties regarding simple waves are
also well established, see \cite{AP87, Anile89}.

For the special case in which the component of the magnetic field
normal to a zone interface vanishes, a degeneracy occurs where
tangential, Alfv\'en and slow waves all propagate at the speed of the
fluid and the solution simplifies to a three-wave pattern, see 
\cite{Rom05}. 

The high degree of nonlinearity inherent to the RMHD equations 
makes seeking for an exact solution prohibitive in terms
of computational costs and efficiency.
For this reasons, approximate methods of solution are preferred instead.

\section{The HLLD Approximate Riemann Solver}
\label{sec:solver}
%
%
%

\begin{figure}\begin{center}
 \includegraphics[width=0.5\textwidth]{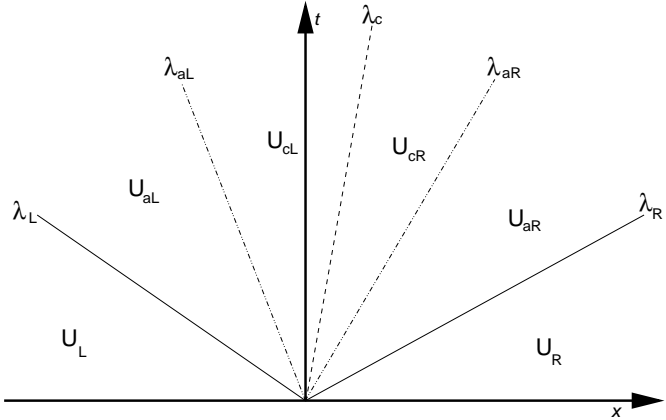}
 \caption{Approximate structure of the Riemann fan introduced by 
          the HLLD solver. The initial states $\vec{U}_L$ and $\vec{U}_R$ 
          are connected to each other through a set of five waves representing, clockwise,
          a fast shock $\lambda_L$, a rotational discontinuity $\lambda_{aL}$,
          a contact wave $\lambda_c$, a rotational discontinuity $\lambda_{aR}$ and 
          a fast shock $\lambda_R$. The outermost states, $\vec{U}_L$ and 
          $\vec{U}_R$ are given as input to the problem, whereas the others
          must be determined consistently solving the Rankine-Hugoniot jump
          conditions.}
 \label{fig:fan}
\end{center}\end{figure}
Without loss of generality, we place the initial discontinuity at
$x = 0$ and set $t^n = 0$.

Following MK, we make the assumption that the Riemann fan 
can be divided by $5$ waves: two outermost
fast shocks, $\lambda_R$ and $\lambda_L$, enclosing two rotational
discontinuities, $\lambda_{aL}$ and $\lambda_{aR}$, separated by
the entropy (or contact) mode with speed $\lambda_c$.
Note that slow modes are not considered in the solution.
The five waves divide the $x-t$ plane into the six regions shown 
in Fig \ref{fig:fan}, corresponding (from left to right) to the $6$
states $\vec{U}_\alpha$ with $\alpha = L, aL, cL, cR, aR, R$.

The outermost states ($\vec{U}_L$ and $\vec{U}_R$)  are given as 
input to the problem, while the remaining ones have to be determined.
In the typical approach used to construct HLL-based solvers,
the outermost velocities $\lambda_L$ and $\lambda_R$ are also provided as
estimates from the input left and right states.
As in MB, we choose to use the simple Davis estimate \citep{Davis88}.

Across any given wave $\lambda$, states and fluxes must satisfy the 
jump conditions
\begin{equation}\label{eq:jump}
 \Big[\lambda\vec{U} - \vec{F}\Big]_{\lambda} \equiv 
 \Big(\lambda\vec{U} - \vec{F}\Big)_+ - \Big(\lambda\vec{U} - \vec{F}\Big)_- = 0\,,
\end{equation}
where $+$ and $-$ identify, respectively, the state immediately 
ahead or behind the wave front. 
Note that for consistency with the integral 
form of the conservation law over the rectangle  
$[\lambda_L\Delta t,  \lambda_R\Delta t]\times[0, \Delta t]$ one has, 
in general, $\vec{F}_{\alpha} \neq \vec{F}(\vec{U}_\alpha)$, except of course
for $\alpha = L$ or $\alpha = R$.

Across the fast waves, we will make frequent use of 
\begin{equation}\label{eq:R}
 \vec{R}_L = \lambda_L\vec{U}_L - \vec{F}_L\,,\quad
 \vec{R}_R = \lambda_R\vec{U}_R - \vec{F}_R\,,\quad
\end{equation}
which are known vectors readily obtained from the left and right input states.
A particular component of $\vec{R}$ is selected by mean of a subscript, e.g., 
$R_D$ is the density component of $\vec{R}$. 

A consistent solution to the problem has to satisfy the $7$ nonlinear 
relations implied by Eq. (\ref{eq:jump}) for each of the $5$ waves considered, 
thus giving a total of $35$ equations.
Moreover, physically relevant solutions must fulfill a number of requirements 
in order to reflect the characteristic nature of the considered waves.
For this reason, across the contact mode, we demand that velocity, magnetic field 
and total pressure be continuous:
\begin{equation}\label{eq:contact_conditions}
   \Big[\vec{v}\Big]_{\lambda_c}  = 
   \Big[\vec{B}\Big]_{\lambda_c}  = \vec{0} \,,\quad
   \Big[p\Big]_{\lambda_c} =  0 \,,
\end{equation}
and require that $\lambda_c\equiv v^x_c$, i.e., that the contact wave moves 
at the speed of the fluid. However, density, energy and total enthalpy 
may be discontinuous. 
On the other hand, through the rotational waves $\lambda_{aL}$ and $\lambda_{aR}$, scalar
quantities such as total pressure and enthalpy are invariant whereas 
all vector components (except for $B^x$) experience jumps.   

%
%
%
             

Since slow magnetosonic waves are not considered, we 
naturally conclude that only the total pressure 
remains constant throughout the fan, contrary to  
Newtonian MHD, where also the velocity normal to the 
interface ($v^x$) is left unchanged across the waves.
This is an obvious consequence of the different nature of 
relativistic Alfv\`en waves across which vector fields like $u^\mu$ 
and $b^\mu$ trace ellipses rather than circles. 
As a consequence, the normal component of the velocity, $v^x$, is 
no longer invariant in RMHD but experiences a jump.
These considerations along with the higher level of complexity of the 
relativistic equations makes the extension of the multi-state HLL solver
to RMHD considerably more elaborate.

Our strategy of solution is briefly summarized. For each state we 
introduce a set of $8$ \emph{independent} unknowns: 
${\cal P}=\left\{D, v^x, v^y, v^z, B^y, B^z, w, p\right\}$ and 
write conservative variables and fluxes given by Eq. (\ref{eq:UandF}) as
\begin{equation}\label{eq:state}
 \vec{U}_\alpha = \left(\begin{array}{c} 
  D                      \\ \noalign{\medskip}
  w\gamma^2v^k - b^0b^k  \\ \noalign{\medskip} 
  w\gamma^2 - p -b^0b^0  \\ \noalign{\medskip}
  B^k\end{array}\right)_\alpha
  \,,
\end{equation}
\begin{equation}\label{eq:flux}
 \vec{F}_\alpha = \left(\begin{array}{c}
  D v^x \\ \noalign{\medskip}
  w\gamma^2v^kv^x - b^kb^x + p\delta_{kx} \\ \noalign{\medskip}
  w\gamma^2v^x - b^0b^x                   \\ \noalign{\medskip}
  B^kv^x - B^xv^k\end{array}\right)_\alpha\,,
\end{equation}
where $k=x,y,z$ labels the vector component, $\alpha$ is the state
and $b^\mu$ is computed directly from (\ref{eq:bmu}).
We proceed by solving, as function of the total pressure $p$, 
the jump conditions (\ref{eq:jump}) across the outermost waves 
$\lambda_L$ and $\lambda_R$.
By requiring that total pressure and Alfv\`en velocity do not change across 
each rotational modes, we find a set of invariant
quantities across $\lambda_{aL}$ and $\lambda_{aR}$. 
Using these invariants, we express states and fluxes on either
side of the contact mode ($\alpha = cL, cR$) in terms of the 
total pressure unknown only.
Imposing continuity of normal velocity, $v^x_{cL}(p) = v^x_{cR}(p)$, leads 
to a nonlinear scalar equation in $p$, whose zero gives the desired solution. 

Once $p$ has been found to some relative accuracy (typically $10^{-6}$), 
the full solution to the problem can be written as
\begin{equation}
 \vec{f} = \left\{\begin{array}{lll}
  \vec{F}_L    & {\rm if} &\lambda_L    > 0                \\ \noalign{\medskip}
  \vec{F}_{aL} & {\rm if} &\lambda_L    < 0 < \lambda_{aL} \\ \noalign{\medskip}
  \vec{F}_{aL} + \lambda_c\left(\vec{U}_{cL} - \vec{U}_{aL}\right) 
               & {\rm if}&\lambda_{aL} < 0 < \lambda_c    \\ \noalign{\medskip}
  \vec{F}_{aR} + \lambda_c\left(\vec{U}_{cR} - \vec{U}_{aR} \right)
               & {\rm if} &\lambda_c    < 0 < \lambda_{aR} \\ \noalign{\medskip}
  \vec{F}_{aR} & {\rm if} &\lambda_{aR} < 0 < \lambda_R    \\ \noalign{\medskip}
  \vec{F}_R    & {\rm if} &\lambda_R    < 0 \\ \noalign{\medskip}
\end{array}\right.
\end{equation}
where $\vec{U}_{aL}, \vec{U}_{aR}$ are computed in \S\ref{sec:fast}, 
$\vec{U}_{cL}, \vec{U}_{cR}$ in \S\ref{sec:contact}
and $\vec{F}_{a} = \vec{F} + \lambda_{a}
(\vec{U}_{a} - \vec{U})$ (for $a=aL$ or $a=aR$) follow from the jump conditions.
The wave speeds $\lambda_{aL},\lambda_{aR}$ and $\lambda_c$ are computed
during the solution process.

Here and in what follows we adopt the convention that single subscripts 
like $a$ (or $c$) refers indifferently to $aL, aR$ (or $cL, cR$).
Thus an expression like $w_c = w_a$ means $w_{cL} = w_{aL}$ and
$w_{cR} = w_{aR}$.

\subsection{Jump Conditions Across the Fast Waves}
\label{sec:fast}
%
%
%

We start by explicitly writing the jump conditions across the outermost 
fast waves:
\begin{eqnarray}
 \label{eq:jumpD}
  \left(\lambda - v^x\right)D & = &   R_D \,, \\ \noalign{\medskip}
\label{eq:jumpmk}
  \left(\lambda - v^x\right)w\gamma^2v^k + b^k\left(b^x - \lambda b^0\right) - p\delta_{kx} & = & R_{m^k}  \,,
\\ \noalign{\medskip}
\label{eq:jumpE}
  (\lambda - v^x)w\gamma^2 - \lambda p   + b^0\left(b^x   - \lambda b^0\right) &= &R_E  \,,
\\ \noalign{\medskip}
\label{eq:jumpBk}
  \left(\lambda - v^x\right)B^k + B^x v^k & = & R_{B^k} \,,
\end{eqnarray}
where, to avoid cluttered notations, we omit in this section
the $\alpha=aL$ (when $\lambda = \lambda_L$) or
$\alpha=aR$ (when $\lambda=\lambda_R$) 
index from the quantities appearing on the left hand side.
Similarly, the $R$'s appearing on the right hand sides of equations 
(\ref{eq:jumpD})--(\ref{eq:jumpBk}) are understood as the components of the 
vector $\vec{R}_L$ (when $\lambda=\lambda_L$) or $\vec{R}_R$ (when 
$\lambda=\lambda_R$), defined by Eq. (\ref{eq:R}).

The jump conditions of Faraday's law allow to express the magnetic field as a 
function of velocities alone, 
\begin{equation}\label{eq:Bv}
  B^k = \frac{R_{B^k} - B^xv^k}{\lambda - v^x} \,\qquad\textrm{for}\qquad 
  k=y,z\,.
\end{equation}
The energy and momentum equations can be combined together to provide an explicit
functional relation between the three components of velocity and the total pressure 
$p$. To this purpose, we first multiply
the energy equation (\ref{eq:jumpE}) times $v^k$ and then subtract the resulting
expression from the jump condition for the $k$-th component of momentum, Eq. (\ref{eq:jumpmk}).
Using Eq. (\ref{eq:jumpBk}) to get rid of the $\vec{v}^2$ term, one finds after some
algebra:
\begin{equation}\label{eq:vel}
 B^k\Big(B^x - \vec{R}_B\cdot\vec{v}\Big) - 
 p\Big(\delta_{kx} - \lambda v^k\Big) = R_{m^k} - v^kR_E\,,
\end{equation}
with $B^k$ defined by (\ref{eq:Bv}).
The system can be solved for $v^k$ giving 
\begin{eqnarray}
\label{eq:vxa}
 v^x & = &\frac{B^x\left(AB^x + \lambda C\right) - 
                \left(A+G\right)\left(p + R_{m^x}\right)}{X}\,,
\\ \noalign{\medskip}
\label{eq:vya}
 v^y &=& \frac{QR_{m^y} + R_{B^y}\left[C + B^x\left(\lambda R_{m^x} - R_E\right)\right]}{X}\,,
\\ \noalign{\medskip}
\label{eq:vza}
 v^z &=& \frac{QR_{m^z} + R_{B^z}\left[C + B^x\left(\lambda R_{m^x} - R_E\right)\right]}{X}\,,
\end{eqnarray}
where 
\begin{eqnarray}
 A & = & R_{m^x} - \lambda R_E + p\left(1-\lambda^2\right)  \,,
\\ 
 G & = & R_{B^y}R_{B^y} + R_{B^z}R_{B^z} \,,
\\
 C  & = &  R_{m^y}R_{B^y} + R_{m^z}R_{B^z} \,,
\\ 
 Q &= & - A - G + (B^x)^2\left(1-\lambda^2\right)
\\
 X &=& B^x\left(A\lambda B^x + C\right) - \left(A+G\right)\left(\lambda p + R_{E}\right)\,.
\end{eqnarray}

Once the velocity components are expressed as functions of $p$, 
the magnetic field is readily found from (\ref{eq:Bv}), while the 
total enthalpy can be found using its definition,
$w = (E + p)/\gamma^2 + (\vec{v}\cdot\vec{B})^2$, or 
by subtracting $R_E$ from the inner product $\vec{v}^k\cdot\vec{R}_m$,
giving
\begin{equation}\label{eq:fastw}
w = p + \frac{R_E - \vec{v}\cdot\vec{R}_m}{\lambda - v^x} \,,
\end{equation}
where $\vec{R}_m \equiv \left(R_{m^x},R_{m^y},R_{m^z}\right)$.
Although equivalent, we choose to use this second expression.
Since the $v^k$ are functions of $p$ alone, the total enthalpy $w$
is also a function of the total pressure. 

The remaining conserved quantities in the $\alpha=aL$ or $\alpha=aR$ regions
can be computed once $p$ has been found: 
\begin{eqnarray}\label{eq:fast_DEM}
 D & = & \DS \frac{R_D}{\lambda - v^x} \,,
  \\ \noalign{\medskip}
 E  & = &  \DS \frac{R_E + pv^x - \left(\vec{v}\cdot\vec{B}\right)B^x}{\lambda - v^x}\,,
  \\ \noalign{\medskip}
 m^k & = & \left(E + p\right)v^k - \left(\vec{v}\cdot\vec{B}\right)B^k\,.
\end{eqnarray}
One can verify by direct substitution that the previous equations 
together with the corresponding 
fluxes, Eq. (\ref{eq:flux}), satisfy the jump conditions given by
(\ref{eq:jumpD})--(\ref{eq:jumpBk}).
 

\subsection{Jump Conditions across the Alfv\`en waves}
\label{sec:Alfven}
%
%
%

Across the rotational waves one could, in principle, proceed as for
the outer waves, i.e., by explicitly writing the jump conditions.
However, as we shall see, the treatment greatly 
simplifies if one introduces the four vector
\begin{equation}\label{eq:sigma}
 \sigma^\mu = \eta u^\mu + b^\mu \,, 
 \qquad\mathrm{with}\qquad
  \eta = \pm{\rm sign}(B^x)\sqrt{w}
\end{equation}
where, for reasons that will be clear later, we take 
the plus (minus) sign for the right (left) state.
From $\sigma^\mu$ we define the spatial vector 
$\vec{K}\equiv(K^x,K^y,K^z)$ with components given by
\begin{equation}\label{eq:K}
  K^k \equiv \frac{\sigma^k}{\sigma^0} = v^k + \frac{B^k}{\gamma\sigma^0}\,.
\end{equation}
The vector $\vec{K}$ has some attractive properties, the most remarkable of which
is that the $x$ component coincides with the propagation speed of the Alfv\`en 
wave \citep{Anile89}.
For this reason, we are motivated to define $\lambda_a \equiv K^x_a$, where
the subscript $a$ stands for either the left or right rotational wave (i.e.
$aL$ or $aR$) since we require that both $K^x$ and $p$ are invariant across the 
rotational discontinuity, i.e., $K^x_c - K^x_a = p_c - p_a = 0$, 
a property certainly shared by the exact solution. 
As we will show, this choice naturally reduces
to the expressions found by MK in the non-relativistic limit.

Indeed, setting $\lambda_a = K^x_a = K^x_c$ and using Eq. (\ref{eq:K}) 
to express $v^k$ as functions of $K^k$, 
the jump conditions simplify to 
\begin{eqnarray}
\label{eq:alfD}
  \left[\frac{D B^x}{\gamma\sigma^0}\right]_{\lambda_a} &= & 0 
 \\ \noalign{\medskip}
\label{eq:alfm}
  \left[\frac{\eta\sigma^k B^x}{\sigma^0} - p\delta_{kx}\right]_{\lambda_a} &=&0
 \\ \noalign{\medskip}
\label{eq:alfE}
  \left[\eta B^x - \frac{\sigma^x}{\sigma^0}p\right]_{\lambda_a} &=&0 
 \\ \noalign{\medskip}
\label{eq:alfB}
  \left[\frac{B^x\sigma^k}{\sigma^0} \right]_{\lambda_a} &=& 0 \,,
\end{eqnarray}
Since also $[p]_{\lambda_a} = 0$, the previous equations further 
imply that (when $B^x\neq 0$) also $D/(\gamma\sigma^0)$, $w$, $K^y$ and
$K^z$ do not change across $\lambda_a$:
\begin{equation}
 \vec{K}_{aL} = \vec{K}_{cL} \equiv \vec{K}_L  \,,\quad
 \eta_{aL} = \eta_{cL} = \eta_L
\end{equation} 
\begin{equation}
 \vec{K}_{aR} = \vec{K}_{cR} \equiv \vec{K}_R  \,,\quad
 \eta_{aR} = \eta_{cR} = \eta_R
\end{equation} 

Being invariant, $\vec{K}$ can be computed from the state lying 
to the left (for $\lambda_{aL}$) or to the right (for $\lambda_{aR}$)
of the discontinuity, thus being a function of the total pressure 
$p$ alone.
Instead of using Eq. (\ref{eq:K}), an alternative and more convenient 
expression may be found by properly replacing $v^k$ with 
$K^k$ in Eq. (\ref{eq:jumpD})--(\ref{eq:jumpBk}). After some algebra one 
finds the simpler expression
\begin{equation}\label{eq:K_simple}
 K^k = \frac{R_{m^k} + p\delta_{kx} + R_{B^k}\eta}
            {\lambda p + R_E + B^x\eta} \,,
\end{equation}
still being a function of the total pressure $p$.

Note that, similarly to its non relativistic limit, we cannot use 
the equations in (\ref{eq:alfD})--(\ref{eq:alfB}) to compute
the solution across the rotational waves, since they do not
provide enough independent relations.
Instead, a solution may be found by considering 
the jump conditions across both rotational discontinuities and 
properly matching them using the conditions at the contact mode.

\subsection{Jump Conditions across the Contact wave}
\label{sec:contact}
%
%
%

At the contact discontinuity (CD) only density and total enthalpy
can be discontinuous, while total pressure, normal and tangential
fields are continuous as expressed by Eq. (\ref{eq:contact_conditions}).

Since the magnetic field is a conserved quantity, one can immediately
use the consistency condition between the innermost waves
$\lambda_{aL}$ and $\lambda_{aR}$ to find $B^k$ across the CD.
Indeed, from
\begin{eqnarray}\label{eq:consistency}
  && \left(\lambda_{c}  - \lambda_{aL}\right)\vec{U}_{cL} +
  \left(\lambda_{aR} - \lambda_{c} \right)\vec{U}_{cR} 
  =    \\ \nonumber
 & = & \lambda_{aR}\vec{U}_{aR} - \lambda_{aL}\vec{U}_{aL} - 
   \vec{F}_{aR} + \vec{F}_{aL}
\end{eqnarray}
one has $B^k_{cL} = B^k_{cR} \equiv B^k_c$, where
\begin{equation}\label{eq:Bc}
 B^k_c =
  \frac{\left[B^k(\lambda - v^x) + B^xv^k\right]_{aR} - 
        \left[B^k(\lambda - v^x) + B^xv^k\right]_{aL}}
       {\lambda_{aR} - \lambda_{aL}} \,.
%
%
\end{equation}
Since quantities in the $aL$ and $aR$ regions are given in terms 
of the $p$ unknown, Eq. (\ref{eq:Bc}) are also functions of $p$
alone.

At this point, we take advantage of the fact that 
$\sigma^\mu u_\mu = -\eta$ to replace
$\gamma\sigma^0$ with $\eta/(1- \vec{K}\cdot\vec{v})$ 
and then rewrite (\ref{eq:K}) as
\begin{equation}\label{eq:KvB}
 K^k = v^k + \frac{B^k}{\eta}\left(1-\vec{K}\cdot\vec{v}\right) \,\qquad 
\textrm{for} \qquad k=x,y,z\,.
\end{equation}
The previous equations form a linear system in the velocity components
$v^k$ and can be easily inverted to the left and to the right of the CD 
to yield
\begin{equation}\label{eq:vc}
 v^k = K^k - 
 \frac{B^k(1 - \vec{K}^2)}{\eta - \vec{K}\cdot{\vec{B}}} \,
\qquad\textrm{for}\qquad k = x,y,z \,.
\end{equation}
which also depend on the total pressure variable only, 
with $w$ and $K^k$ being given by (\ref{eq:fastw}) and
(\ref{eq:K_simple}) and the $B^k_c$'s being computed from Eq. (\ref{eq:Bc}).
Imposing continuity of the normal velocity across the CD,
$v^x_{cL} - v^x_{cR} = 0$, results in 
\begin{equation}\label{eq:final}
 \Delta K^x \Big[1 -  B^x\Big(Y_R - Y_L\Big)\Big] = 0 \,,
\end{equation}
where
\begin{equation}
 Y_S(p) = \frac{1-\vec{K}^2_{S}}
               {\eta_{S}\Delta K^x - \vec{K}_{S}\cdot\hat{\vec{B}}_c}\,, 
\qquad S=L,R\,,
\end{equation}
is a function of $p$ only 
and $\hat{\vec{B}}_c \equiv \Delta K^x \vec{B}_c$ is the numerator of (\ref{eq:Bc})
and $\Delta K^x = K^x_{aR} - K^x_{aL}$.
Equation (\ref{eq:final}) is a nonlinear function in $p$ and must be solved 
numerically.

Once the iteration process has been completed and $p$ has been found to 
some level of accuracy, the remaining conserved variables 
to the left and to the right of the CD are computed from the jump conditions 
across $\lambda_{aL}$ and $\lambda_{aR}$ and the definition of the flux, 
Eq. (\ref{eq:flux}).
Specifically one has, for 
$\{c=cL, a=aL\}$ or $\{c=cR, a=aR\}$,

\begin{eqnarray}
 D_{c} & = & D_{a}\frac{\lambda_{a} - v^x_{a}}{\lambda_{a} - v^x_{c}} \,, \\ \noalign{\medskip}
 E_{c} & = & \frac{\lambda_{a} E_{a} - m^x_{a} + pv^x_{c} - 
             \left(\vec{v}_{c}\cdot\vec{B}_{c}\right)B^x}
            {\lambda_{a} - v^x_{c}} \,, \\ \noalign{\medskip}
 m^k_{c} & = & (E_{c} + p)v^k_{c} - 
                \left(\vec{v}_{c}\cdot\vec{B}_{c}\right)B^k_{c}\,.
\end{eqnarray}
This concludes the derivation of our Riemann solver.

\subsection{Full Solution}
%
%
%
%

In the previous sections we have shown that the whole set of 
jump conditions can be brought down to the solution of a single nonlinear 
equation, given by (\ref{eq:final}), in the total pressure variable $p$.
In the particular case of vanishing normal component of the magnetic
field, i.e. $B_x\to 0$, this equation can be solved exactly
as discussed in \S\ref{sec:Bx0}.

For the more general case, the solution has to be found numerically using
an iterative method where, starting from an initial guess $p^{(0)}$, 
each iteration consists of the following steps:
\begin{itemize}
 \item given a new guess value $p^{(k)}$ to the total pressure, 
       start from Eq. (\ref{eq:vxa})--(\ref{eq:vza})
       to express $\vec{v}_{aL}$ and $\vec{v}_{aR}$ as functions
       of the total pressure. Also, express magnetic fields $\vec{B}_{aL}$,  
       $\vec{B}_{aR}$ and total enthalpies $w_L$, $w_R$ using
       Eq. (\ref{eq:Bv}) and Eq. (\ref{eq:fastw}), respectively.
 \item Compute $\vec{K}_{aL}$ and $\vec{K}_{aR}$ using Eq. (\ref{eq:K_simple})
       and the transverse components of $\vec{B}_c$ using
       Eq. (\ref{eq:Bc}). 
 \item Use Eq. (\ref{eq:final}) to find the next improved iteration value.
\end{itemize}
For the sake of assessing the validity of our new solver, 
we choose the secant method as our root-finding algorithm.
The initial guess is provided using the following prescription:
\begin{equation}\label{eq:guess}
 p^{(0)} = \left\{\begin{array}{cl}
   p_0  & \quad\mathrm{when} \quad  (B^x)^2/p^{\rm hll} < 0.1  \,,
   \\  \noalign{\medskip}
   p^{\rm hll} & \quad\mathrm{otherwise}  \,,
 \end{array}\right.
\end{equation}
where $p^{\rm hll}$ is the total pressure computed from the HLL 
average state whereas $p_0$ is the solution in the $B^x=0$ limiting
case.
Extensive numerical testing has shown that the total pressure 
$p^{\rm hll}$ computed from the HLL average state provides, in most cases, 
a sufficiently close guess to the correct physical solution, 
so that no more than $5-6$ iterations (for zones with steep gradients)
were required to achieve a relative accuracy of $10^{-6}$.

The computational cost depends on the simulation setting 
since the average number of iterations can vary from one problem to 
another. However, based on the results presented in \S\ref{sec:num},
we have found that HLLD was at most a factor of $\sim 2$ slower than HLL.

For a solution to be physically consistent and well-behaved, 
we demand that 
\begin{equation}\label{eq:conditions}
 \left\{\begin{array}{lll}
 w_{L}  > p \,, & v^x_{aL} > \lambda_{L} \,, & v^x_{cL} > \lambda_{aL} \,, \\ \noalign{\medskip}
 w_{R}  > p \,, & v^x_{aR} < \lambda_{R} \,, & v^x_{cR} < \lambda_{aR} \,,
\end{array}\right.
\end{equation}
hold simultaneously. These conditions guarantee positivity 
of density and that the correct eigenvalue ordering is always
respected.
We warn the reader that equation (\ref{eq:final}) may have, in general, 
more than one solution and that the conditions given by (\ref{eq:conditions})
may actually prove helpful in selecting the correct one.
However, the intrinsic nonlinear complexity of the RMHD equations 
makes rather arduous and challenging to prove, \emph{a priori}, 
both the existence and the uniqueness of a physically relevant solution, 
in the sense provided by (\ref{eq:conditions}). 
On the contrary, we encountered sporadic situations where 
none of the zeroes of Eq. (\ref{eq:final}) is physically admissible.
Fortunately, these situations turn out to be rare eventualities caused 
either by a large jump between left and right states  (as at the
beginning of integration) or by under- or 
over- estimating the propagation speeds of the outermost fast waves,
$\lambda_L$ and $\lambda_R$.
The latter conclusion is supported by the fact that, enlarging 
one or both wave speeds, led to a perfectly smooth and unique solution. 

Therefore, we propose a safety mechanism whereby we switch to the
simpler HLL Riemann solver whenever at least one or more
of the conditions in (\ref{eq:conditions}) is not fulfilled.
From several numerical tests, including the ones shown here, we found 
the occurrence of these anomalies to be limited to 
few zones of the computational domain, usually 
less than $0.1\%$ in the tests presented here.

We conclude this section by noting that other more sophisticated algorithms 
may in principle be sought. 
One could, for instance, provide a better guess to the
outer wave-speeds $\lambda_L$ and $\lambda_R$ or even modify them 
accordingly until a solution is guaranteed to exist.
Another, perhaps more useful, possibility is to bracket the solution 
inside a closed interval $[p_{\min}, p_{\max}]$ where $p_{\min}$ and 
$p_{\max}$ may be found from the conditions (\ref{eq:conditions}).
Using an alternative root finder, such as Ridder \citep{NumRec}, guarantees that the 
solution never jump outside the interval.
However, due to the small number of failures usually encountered,
we do not think these alternatives could lead to a significant gain 
in accuracy.

\subsubsection{Zero normal field limit}
\label{sec:Bx0}
%
%

In the limit $B^x\to 0$ a degeneracy occurs where the Alfv\`en
(and slow) waves propagate at the speed of the contact mode
which thus becomes a tangential discontinuity.
Across this degenerate front, only normal velocity and total pressure
remain continuous, whereas tangential vector fields are subject to jumps.

This case does not pose any serious difficulty in our derivation
and can be solved exactly.
Indeed, by setting $B^x=0$ in Eq. (\ref{eq:K_simple}) and (\ref{eq:final}), 
one immediately finds that $K^x_R = K^x_L = v^x_c$ leading
to the following quadratic equation for $p$:
\begin{equation}\label{eq:Bx0}
p^2 + \left(E^{\rm hll} - F^{\rm hll}_{m^x}\right)p 
    + m^{x,\rm hll}F^{\rm hll}_E - F^{\rm hll}_{m^x}E^{\rm hll} = 0 \,,
\end{equation}
where the superscript ``hll" refers to the HLL average state 
or flux given by Eq. (28) or (31) of MB.
We note that equation (\ref{eq:Bx0}) coincides with the derivation
given by MB \citep[see also][]{MMB05} in the same degenerate case and
the positive root gives the correct physical solution.	            
The intermediate states, $\vec{U}_{cL}$ and $\vec{U}_{cR}$, loose
their physical meaning as $B^x\to 0$ but they never
enter the solution since, as $\lambda_{aL},\lambda_{aR}\to\lambda_c$, only
$\vec{U}_{aL}$ and $\vec{U}_{aR}$ will have a nonzero finite
width, see Fig. \ref{fig:fan}.

Given the initial guess, Eq. (\ref{eq:guess}),
our proposed approach does not have to deal 
separately with the $B^x \neq 0$ and $B^x =0$ cases 
\cite[as in MB and][]{HJ07} and thus
solves the issue raised by MB.

\subsubsection{Newtonian Limit}
%
%
%
%
%

We now show that our derivation reduces to the HLLD Riemann solver
found by MK under the appropriate non-relativistic limit.
We begin by noticing that, for $\vec{v}/c \to 0$, the velocity and induction
four-vectors reduce to $u^\mu\to (1, v^k)$ and $b^\mu\to (0,B^k)$, respectively.
Also, note that $w_g,w\to \rho$ in the non-relativistic limit 
so that 
\begin{equation}
 K^k \to v^k + s\frac{B^k}{\sqrt{\rho}} \,,
\end{equation}
and thus $v^x$ cannot change across $\lambda_a$.
Replacing (\ref{eq:jumpD})-(\ref{eq:jumpmk}) with their
non-relativistic expressions and demanding $v^x_{a} = v^x_{c}$ 
gives, in our notations, the following expressions:
\begin{eqnarray}
v^x_a & = & \frac{R_{R,m^x} - R_{L,m^x}}{R_{R,D} - R_{L,D}}  \,,  \\
p     & = & (B^x)^2 - \frac{R_{L,m^x}R_{R_{R,D}} - R_{R,m^x}R_{R_{L,D}}}
                           {R_{R,D} - R_{L,D}} \,,
\end{eqnarray}
which can be shown to be identical to Eqns (38) and (41) of MK.
With little algebra, one can also show that the remaining variables
in the $aL$ and $aR$ regions reduce to the corresponding non-relativistic
expressions of MK.
Similarly, the jump across the rotational waves are solved exactly
in the same fashion, that is, by solving the integral conservation laws
over the Riemann fan. For instance, Eq. (\ref{eq:Bc}) reduces to 
equation (61) and (62) of MK.
These results should not be surprising since, our set of parameters to write
conserved variables and fluxes is identical to the one used by 
MK. The only exception is the energy, which is actually written 
in terms of the total enthalpy.

\section{Numerical Tests}
\label{sec:num}
%
%
%

\setlength\tabcolsep{1.5mm}
\begin{table*}
\centering
\begin{tabular}{cc cccccccc cc}\hline
Test   & State & $\rho$ & $p_g$ & $v^x$ & $v^y$ & $v^z$ & $B^x$ & $B^y$ & $B^z$ & Time & Zones\\ \hline\hline
\multirow{2}{*}{Contact Wave}  
  & L  & $10$ & $1$ & $0$  & $0.7$ & $0.2$ & $5$ & $1$ & $0.5$ & \multirow{2}{*}{1} & \multirow{2}{*}{40}\\ 
  & R  &  $1$ & $1$ & $0$  & $0.7$ & $0.2$ & $5$ & $1$ & $0.5$ &  &\\ \hline
\multirow{2}{*}{Rotational Wave}  & 
    L  & $1$ & $1$ & $0.4$      & $-0.3$      & $0.5$      & $2.4$ & $1$    & $-1.6$ & \multirow{2}{*}{1} & \multirow{2}{*}{40}\\ 
  & R  & $1$ & $1$ & $0.377347$ & $-0.482389$ & $0.424190$ & $2.4$ & $-0.1$ & $-2.178213$ &   &\\ \hline
\multirow{2}{*}{Shock Tube 1}  & 
    L  & $1$     & $1$   & $0$  & $0$ & $0$ & $0.5$ & $ 1$ & $0$ & \multirow{2}{*}{0.4} & \multirow{2}{*}{400}\\ 
  & R  & $0.125$ & $0.1$ & $0$  & $0$ & $0$ & $0.5$ & $-1$ & $0$ &   &\\ \hline
\multirow{2}{*}{Shock Tube 2}  & 
    L  & $1.08$ & $0.95$ & $0.4$   & $0.3$  & $0.2$ & $2$ & $0.3$  & $0.3$ & \multirow{2}{*}{0.55} & \multirow{2}{*}{800}\\ 
  & R  & $1$    & $1$    & $-0.45$ & $-0.2$ & $0.2$ & $2$ & $-0.7$ & $0.5$ &  &\\ \hline
\multirow{2}{*}{Shock Tube 3}  & 
    L  & $1$ & $0.1$ & $ 0.999$ & $0$ & $0$ & $10$ & $ 7$  & $7$  & \multirow{2}{*}{0.4} & \multirow{2}{*}{400}\\ 
  & R  & $1$ & $0.1$ & $-0.999$ & $0$ & $0$ & $10$ & $-7$  & $-7$ &   &\\ \hline
\multirow{2}{*}{Shock Tube 4}  & 
    L  & $1$   & $5$   & $0$ & $0.3$ & $0.4$ & $1$ & $6$ & $2$ & \multirow{2}{*}{0.5} & \multirow{2}{*}{800}\\ 
  & R  & $0.9$ & $5.3$ & $0$ & $0$   & $0$   & $1$ & $5$ & $2$ &   &\\ \hline
\end{tabular}
\label{tab:ic}
\caption{Initial conditions for the test problems discussed in the text. The last two columns
         give, respectively, the final integration time and the number of computational zones 
         used in the computation.}
\end{table*}

We now evaluate, in \S\ref{sec:test1d}, the accuracy of the proposed 
HLLD Riemann solver by means of selected one dimensional shock tube problems.
Applications of the solver to multi-dimensional problems of 
astrophysical relevance are presented in \S\ref{sec:test23d}. 

\subsection{One Dimensional Shock Tubes}
\label{sec:test1d}
%
%
%

The initial condition is given by Eq. (\ref{eq:riemann})
with left and right states defined by the primitive variables listed
in Table 1.
The computational domain is chosen to be the interval $[0,1]$ and 
the discontinuity is placed at $x=0.5$. The resolution $N_x$ and final 
integration time can be found in the last two columns of 
Table 1.
Unless otherwise stated, we employ the constant $\Gamma-$ law with 
$\Gamma=5/3$.
The RMHD equations are solved using the first-order accurate scheme
(\ref{eq:1st_ord}) with a CFL number of $0.8$.

Numerical results are compared to the HLLC Riemann solver of MB and
to the simpler HLL scheme and the accuracy is quantified by computing discrete 
errors in L-1 norm:
\begin{equation}\label{eq:error}
 \epsilon_{\rm L1} = \sum_{i=1}^{i=N_x} 
                      \left|q^{\rm ref}_i - q_i\right|\Delta x_i \,,
\end{equation}
where $q_i$ is the first-order numerical solution 
(density or magnetic field), $q^{\rm ref}_i$ is the reference solution 
at $x_i$ and $\Delta x_i$ is the mesh spacing.
For tests $1,2,4$ we obtained a reference solution using the second-order 
scheme of MB on $3200$ zones and adaptive mesh refinement with 
$6$ levels of refinement (equivalent resolution $204,800$ grid points). 
Grid adaptivity in one dimension has been incorporated in the 
PLUTO code using a block-structured grid approach following \cite{BC89}.
For test $3$, we use the exact numerical solution available from 
\cite{GR06}. Errors (in percent) are shown in Fig. \ref{fig:error}.

\subsubsection{Exact Resolution of Contact and Alfv\`en Discontinuities}
%
%
%
%
\begin{figure}\begin{center}
 \includegraphics[width=0.5\textwidth]{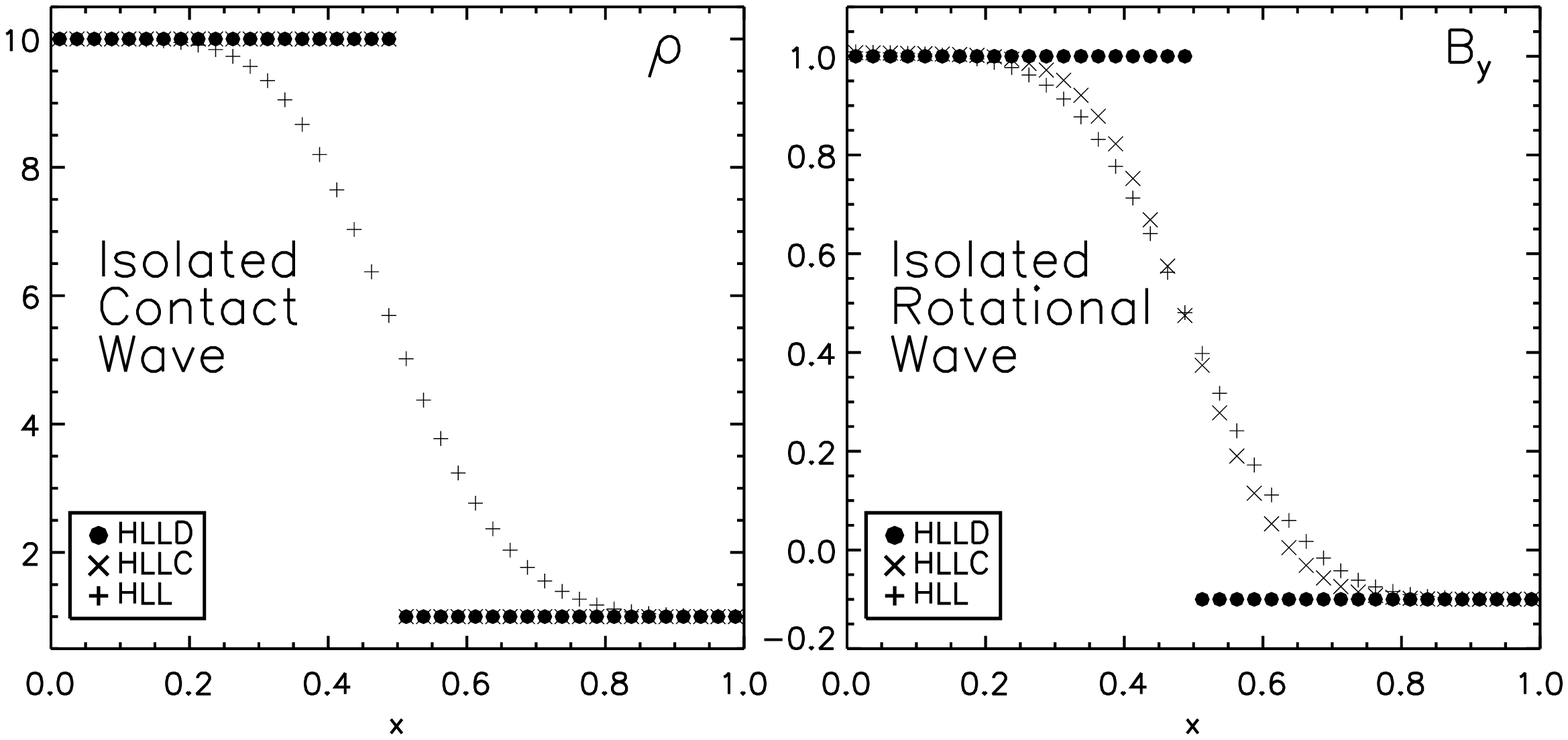}
 \caption{Results for the isolated contact (left panel)
          and rotational (right panel) waves at $t=1$.
          Density and $y$ component of magnetic field are shown, respectively.
          The different symbols show results computed with the new HLLD solver 
          (filled circles), the HLLC solver (crosses) and
          the simpler HLL solver (plus signs). Note that only HLLD is able
          to capture exactly both discontinuities by keeping
          them perfectly sharp without producing any grid diffusion effect.
          HLLC can capture the contact wave but
          not the rotational discontinuity, whereas HLL spreads
          both of them on several grid zones.}
 \label{fig:isolated}
\end{center}\end{figure}
We now show that our HLLD solver can capture \emph{exactly} isolated
contact and rotational discontinuities.
The initial conditions are listed at the beginning of Table 1.

In the case of an isolated stationary contact wave, only density 
is discontinuous across the interface.
The left panel in Fig. \ref{fig:isolated} shows the results
at $t=1$ computed with the HLLD, HLLC and HLL solvers:
as expected our HLLD produces no smearing of the discontinuity
(as does HLLC).
On the contrary, the initial jump broadens over several grid
zone when employing the HLL scheme.

Across a rotational discontinuity, scalar quantities such as 
proper density, pressure and total enthalpy are invariant but
vector fields experience jumps.
The left and right states on either side of an exact rotational 
discontinuity can be found using the procedure outlined in the Appendix.
The right panel in Fig. \ref{fig:isolated} shows that only HLLD
can successfully keep a sharp resolution of the discontinuity, whereas
both HLLC and HLL spread the jump over several grid points because of the 
larger numerical viscosity.

\subsubsection{Shock Tube 1}
%
%
%
%

\begin{figure}\begin{center}
 \includegraphics[width=0.5\textwidth]{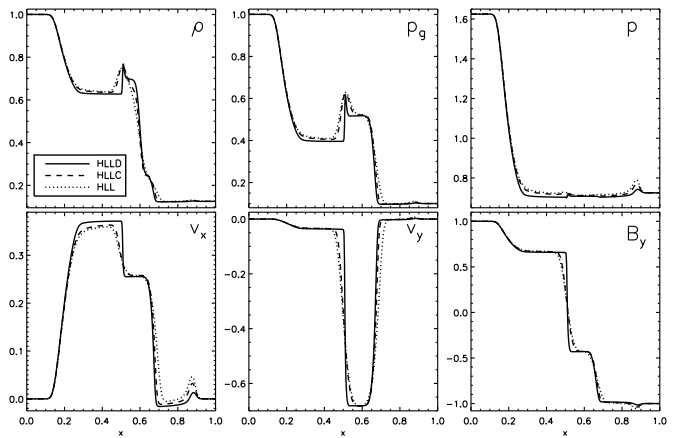}
 \caption{Relativistic Brio-Wu shock tube test at $t=0.4$.
          Computations are carried on $400$ zones using the HLLD (solid line),
          HLLC (dashed line) and HLL (dotted line) Riemann solver, respectively.
          The top panel shows, from left to right, the rest mass 
          density, gas pressure,
          total pressure. The bottom panel shows the $x$ and $y$ components of velocity 
          and the $y$ component of magnetic field.}
 \label{fig:st1}
\end{center}\end{figure}
\begin{figure}\begin{center}
 \includegraphics[width=0.5\textwidth]{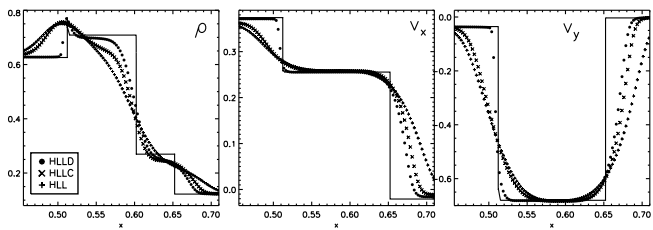}
 \caption{Enlargement of the central region of Fig. \ref{fig:st1}. 
          Density and the two components of velocity are shown in the 
          left, central and right panels, respectively. Diamonds, crosses 
          and plus signs are used for the HLLD, HLLC and HLL Riemann solver,
          respectively.}
 \label{fig:st1_close}
\end{center}\end{figure}

The first shock tube test is a relativistic extension of the 
Brio Wu magnetic shock tube \citep{BW88} and has also been
considered by \cite{Balsara01, dZBL03} and in MB. 
The specific heat ratio is $\Gamma = 2$.
The initial discontinuity breaks into a left-going fast
rarefaction wave, a left-going compound wave, a contact discontinuity,
a right-going slow shock and a right-going fast rarefaction wave.
Rotational discontinuities are not generated.

In Figs. \ref{fig:st1}-\ref{fig:st1_close} we plot the results 
obtained with the first-order scheme and compare them with the 
HLLC Riemann solver of MB and the HLL scheme.
Although the resolution across the continuous right-going rarefaction wave
is essentially the same, the HLLD solver offers a considerable improvement in accuracy 
in the structures located in the central region of the plots.
Indeed, Fig. \ref{fig:st1_close} shows an enlargement of the central part of the domain,
where the compound wave (at $x\approx 0.51$), contact ($x\approx 0.6$)
and slow shock ($x\approx 0.68$) are clearly visible.
Besides the steeper profiles of the contact and slow modes, it is 
interesting to notice that the compound wave, composed of a slow
shock adjacent to a slow rarefaction wave, is noticeably better
resolved with the HLLD scheme than with the other two.

These results are supported by the convergence study 
shown in the top left panel of Fig. \ref{fig:error}, demonstrating
that the errors obtained with our new scheme are smaller than those 
obtained with the HLLC and HLL solvers (respectively). 
At the largest resolution employed, for example,
the L-1 norm errors become $\sim 63\%$ and $\sim 49\%$ smaller than the HLL
and HLLC schemes, respectively.

 The CPU times required by the different Riemann solvers on this particular 
 test were found to be scale as
 $t_{\rm hll}:t_{\rm hllc}:t_{\rm hlld} = 1:1.2:1.9$.


\subsubsection{Shock Tube 2}
%
%
%
%

\begin{figure}\begin{center}
 \includegraphics[width=0.5\textwidth]{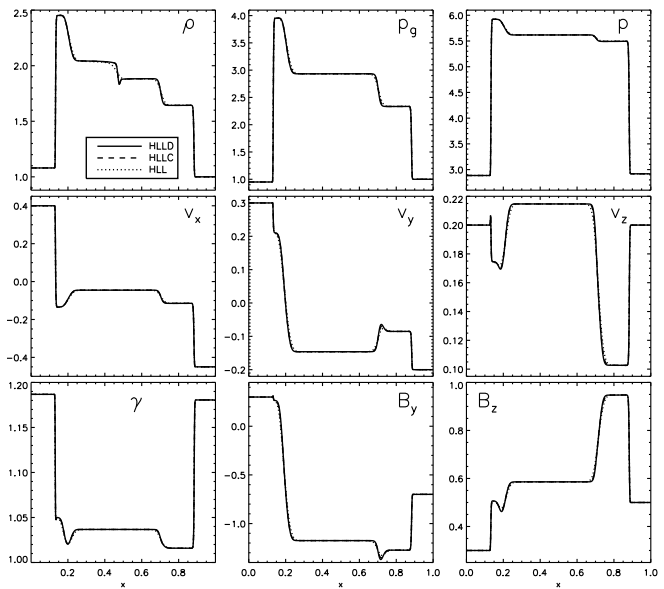}
 \caption{Results for the second shock tube problem at 
          $t=0.55$ on $800$ grid points.
          From left to right, the top panel shows density, gas 
          and total pressure.
          The middle panel shows the three components of velocity, whereas
          in the bottom panel we plot the Lorentz factor and the
          transverse components of magnetic field.
          Solid, dashed and dotted lines are used to identify results
          computed with HLLD, HLLC and HLL, respectively.}
 \label{fig:st2}
\end{center}\end{figure}

\begin{figure}\begin{center}
 \includegraphics[width=0.5\textwidth]{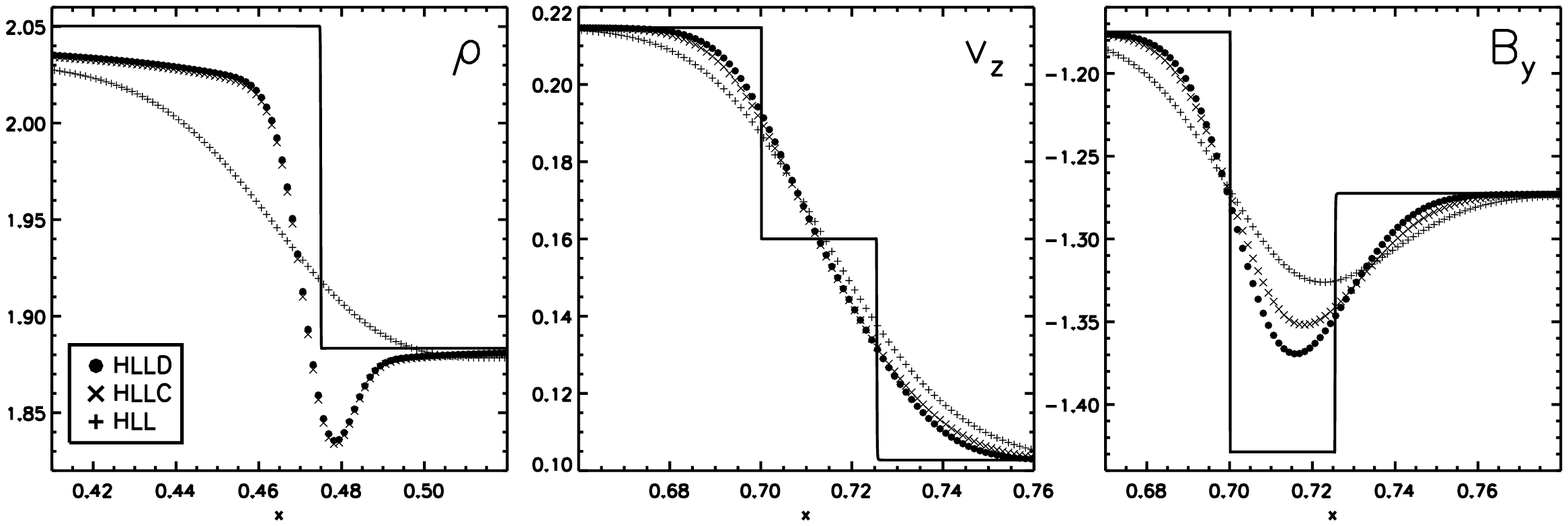}
 \caption{Left panel: enlargement of the central 
          region of Fig. \ref{fig:st2} around the contact wave. 
          Middle and right panels: close-ups of the $z$ component of velocity 
          and $y$ component of magnetic field around the right-going
          slow shock and Alfv\`en discontinuity. 
          Different symbols refer to different Riemann solver, see
          the legend in the left panel.}
 \label{fig:st2_close}
\end{center}\end{figure}

This test has also been considered in \cite{Balsara01} and in MB
and the initial condition comes out as a non-planar Riemann problem
implying that the change in orientation of the transverse magnetic field
across the discontinuity is $\approx 0.55\pi$ (thus different from zero
or $\pi$).

The emerging wave pattern consists of a contact wave (at $x\approx 0.475$)
separating a left-going fast shock ($x\approx 0.13$),
Alfv\`en wave ($x\approx 0.185$) and slow rarefaction ($x\approx 0.19$)
from a slow shock ($x\approx 0.7$), 
Alfv\`en wave ($x\approx 0.725$) and fast shock ($x\approx 0.88$)
heading to the right.

Computations carried out with the $1^{\rm st}$ order accurate scheme are
shown in Fig. \ref{fig:st2} using the HLLD (solid line), HLLC 
(dashed line) and HLL (dotted line). 
The resolution across the outermost fast shocks is essentially the same 
for all Riemann solvers.
Across the entropy mode both HLLD and HLLC attain a sharper representation 
of the discontinuity albeit unphysical undershoots are visible immediately 
ahead of the contact mode.
This is best noticed in the the left panel of Fig. \ref{fig:st2_close},
where an enlargement of the same region is displayed.
 
On the right side of the domain, the slow shock and the rotational
wave propagate quite close to each other and the first-order
scheme can barely distinguish them at a resolution of $800$ zones.
However, a close-up of the two waves (middle and right panel
in Fig. \ref{fig:st2_close}) shows that the proposed scheme is still 
more accurate than HLLC in resolving both fronts.

On the left hand side, the separation between the Alfv\`en and slow 
rarefaction waves turns out to be even smaller and the two modes blur 
into a single wave because of the large numerical viscosity.
This result is not surprising since these features are, in fact,
challenging even for a second-order scheme \citep{Balsara01}. 

Discrete L-1 errors computed using Eq. (\ref{eq:error}) are plotted as 
function of the resolution in the top right panel of Fig. \ref{fig:error}.
For this particular test, HLLD and HLLC produce comparable errors
($\sim 1.22\%$ and $\sim 1.33\%$ at the highest resolution) while HLL
performs worse on contact, slow and Alfv\`en waves resulting in larger
deviations from the reference solution.

The computational costs on $800$ grid zones has found to be
$t_{\rm hll}:t_{\rm hllc}:t_{\rm hlld} = 1:1.1:1.6$.

\subsubsection{Shock Tube 3}
%
%
%
%

\begin{figure}\begin{center}
 \includegraphics[width=0.5\textwidth]{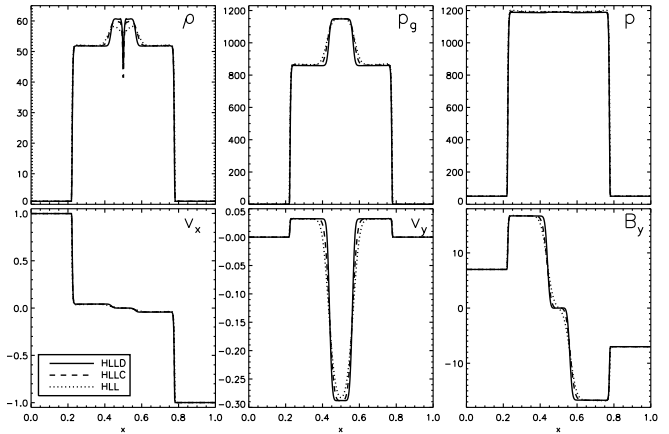}
 \caption{Relativistic collision of two oppositely moving
          streams at $t=0.4$. From top to bottom, left to right, the panels show 
          density $\rho$, gas pressure  $p_g$, total pressure $p$, 
          $x$ and $y$ components of velocity ($v^x$ $v^y$) and
          $y$ component of magnetic field $B^y$. The $z$ components
          have been omitted since they are identical to the $y$ components.
          Solid, dashed and dotted lines refer to computations
          obtained with the HLLD, HLLC and HLL solvers.
          $400$ computational zones were used in the computations.}
 \label{fig:st3}
\end{center}\end{figure}

\begin{figure}\begin{center}
 \includegraphics[width=0.5\textwidth]{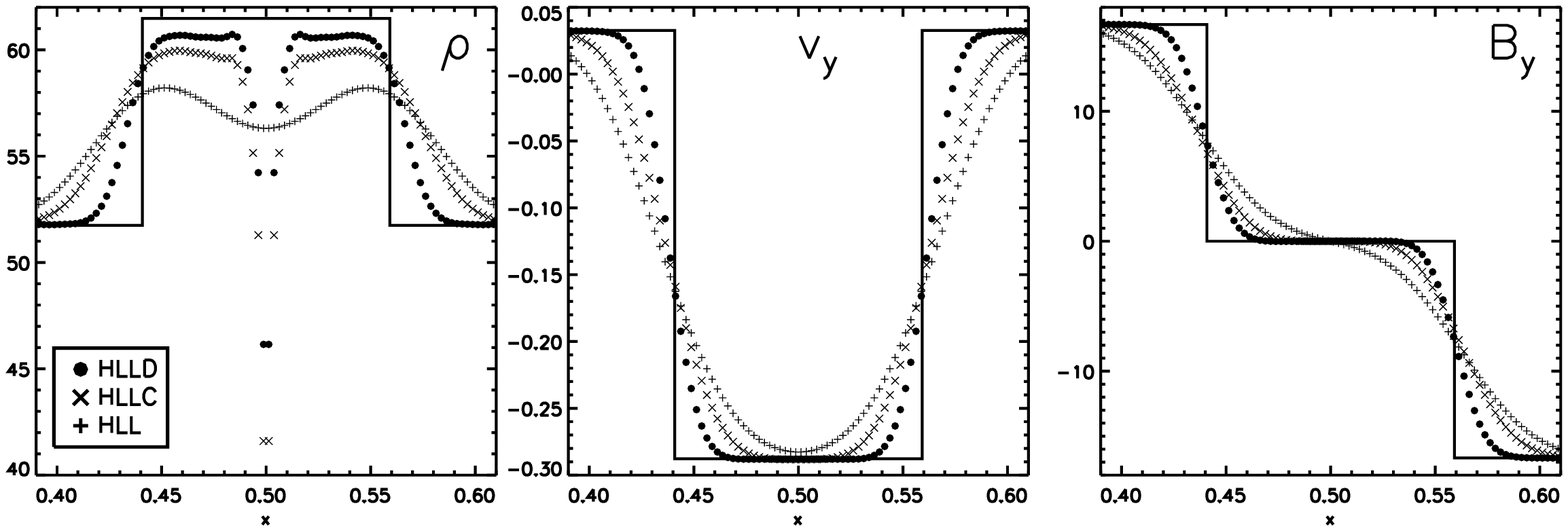}
 \caption{Enlargement of the central region in 
          Fig. \ref{fig:st3}. Filled circles crosses and plus sign have the
          same meaning as in Fig. \ref{fig:st2_close}.
          Note the wall heating problem evident in the density 
          profile (left panel). Central and right panels show
          the transverse field profiles. Clearly the resolution 
          of the slow shocks ($x\approx 0.5\pm 0.07$) improves
          from HLL to HLLC and more from HLLC to HLLD.}
 \label{fig:st3_close}
\end{center}\end{figure}

In this test problem we consider the interaction of two oppositely colliding 
relativistic streams, see also \cite{Balsara01,dZBL03} and MB.

After the initial impact, two strong relativistic fast shocks
propagate outwards symmetrically in opposite direction about the impact 
point, $x=0.5$, see Fig. \ref{fig:st3}. 
Being a co-planar problem (i.e. the initial twist angle between 
magnetic fields is $\pi$) no rotational mode can actually appear.
Two slow shocks delimiting a high pressure constant 
density region in the center follow behind.

Although no contact wave forms, the resolution across the slow shocks
noticeably improves changing from HLL to HLLC and from HLLC to HLLD, 
see Fig. \ref{fig:st3} or the enlargement of the central 
region shown in Fig. \ref{fig:st3_close}. 
The resolution across the outermost fast shocks is essentially the same
for all solvers.

The spurious density undershoot at the center of the grid is a notorious
numerical pathology, known as the wall heating problem,
often encountered in Godunov-type schemes \citep{Noh87, GCM97}.
It consists of an undesired entropy buildup in a few zones around 
the point of symmetry.
Our scheme is obviously no exception as it can be 
inferred by inspecting see Fig. \ref{fig:st3}.
Surprisingly, we notice that error HLLD performs slightly better than HLLC.
The numerical undershoots in density, in fact, are found to be
$\sim 24\%$ (HLLD) and $\sim 32\%$ (HLLC).
The HLL solver is less prone to this pathology most likely because of the
larger numerical diffusion, see the left panel close-up 
of Fig. \ref{fig:st3_close}.

Errors (for $B^y$) are computed using the exact solution available 
from \cite{GR06} which is free from the pathology just discussed.
As shown in the bottom left panel of Fig. \ref{fig:error}, HLLD
performs as the best numerical scheme yielding, at the largest 
resolution employed ($3200$ zones), L-1 norm errors 
of $\sim 18\%$ to be compared to $\sim 32\%$ and $\sim 46\%$ of  
HLLC and HLL, respectively. 

The CPU times for the different solvers on this problem follow
the proportion $t_{\rm hll}:t_{\rm hllc}:t_{\rm hlld} = 1:1.1:1.4$.

\subsubsection{Shock Tube 4}
%
%
%
%

\begin{figure}\begin{center}
 \includegraphics[width=0.5\textwidth]{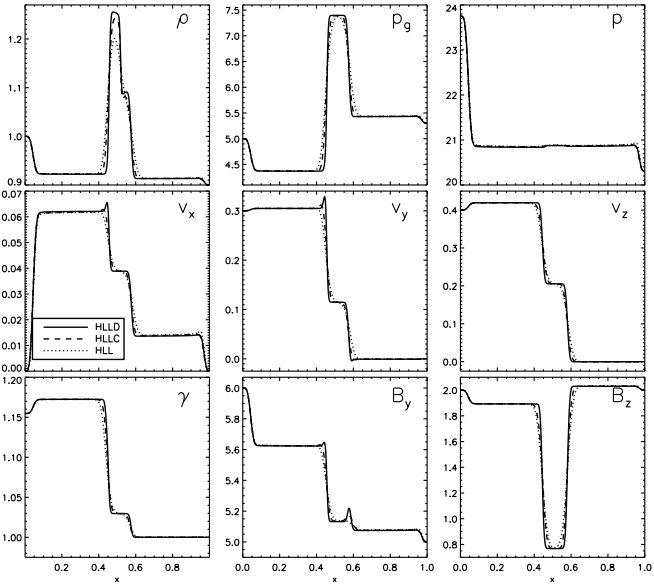}
 \caption{Results for the general Alfv\`en test, problem 4,
          at $t=0.5$ on $800$ computational zones.
          The panels are structured in a way similar to Fig. \ref{fig:st2}.
          Top panel: density, gas pressure and total pressure.
          Mid panel: $x$, $y$ and $z$ velocity components.
          Bottom panel: Lorentz factor $\gamma$ and transverse
          components of magnetic field.}
 \label{fig:st4}
\end{center}\end{figure}

\begin{figure}\begin{center}
 \includegraphics[width=0.5\textwidth]{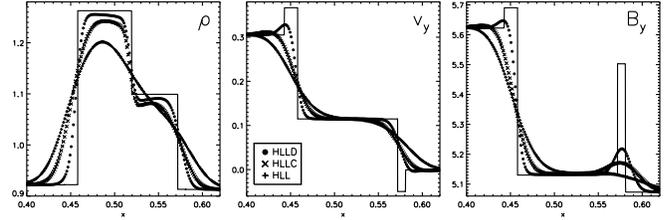}
 \caption{Magnification of the central region
          of Fig. \ref{fig:st4}. The left panel shows the density 
          profile where the two slow shocks and the central 
          contact wave are clearly visible. 
          Central and right panels display the
          $y$ components of velocity and magnetic field. 
          Rotational modes can be most clearly distinguished only 
          with the HLLD solver at $x\approx 0.44$ and $x\approx 0.59$.}
 \label{fig:st4_close}
\end{center}\end{figure}
\begin{figure}\begin{center}
 \includegraphics[width=0.5\textwidth]{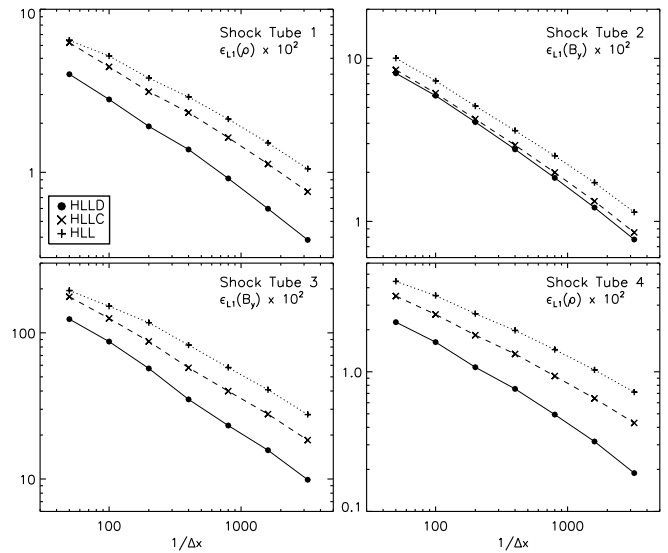}
 \caption{L-1 norm errors (in $10^2$) for the four
          shock tube problems presented in the text as function 
          of the grid resolution.
          The different combinations of lines and symbols 
          refer to HLLD (solid, circles), HLLC (dashed, crosses)
          and HLL (dotted, plus signs).}
 \label{fig:error}
\end{center}\end{figure}

The fourth shock tube test is taken from the ``Generic Alfv\`en" 
test in \cite{GR06}.
The breaking of the initial discontinuous states leads 
to the formation of seven waves.
To the left of the contact discontinuity one has 
a fast rarefaction wave, followed
by a rotational wave and a slow shock.
Traveling to the right of the contact discontinuity, one can find 
a slow shock, an Alfv\`en wave and a fast shock.

We plot, in Fig. \ref{fig:st4}, the results computed with 
the HLLD, HLLC and HLL Riemann solvers at $t=0.5$, when
the outermost waves have almost left the outer boundaries.
The central structure ($0.4\lesssim x\lesssim 0.6$) 
is characterized by slowly moving fronts with the rotational discontinuities
propagating very close to the slow shocks.
At the resolution employed ($800$ zones), the rotational and slow modes 
appear to be visible and distinct only with the HLLD solver, 
whereas they become barely discernible with the HLLC solver
and completely blend into a single wave using the HLL scheme.
This is better shown in the enlargement of $v^y$ and $B^y$ profiles shown 
in Fig. \ref{fig:st4_close}: rotational modes are captured at
$x\approx 0.44$ and $x\approx 0.59$ with the HLLD solver and gradually
disappear when switching to the HLL scheme.

At the contact wave HLLD and HLLC behave similarly but the sharper
resolution attained at the left-going slow shock allows to 
better capture the constant density shell between the two fronts.

Our scheme results in the smallest errors and numerical 
dissipation and exhibits a slightly
faster convergence rate, see the plots in the bottom right panel of 
Fig. \ref{fig:error}.
At low resolution the errors obtained with HLL, HLLC and HLLD
are in the ratio $1:0.75:0.45$ while they become $1:0.6:0.27$ as the
mesh thickens.
Correspondingly, the CPU running times for the three solvers 
at the resolution shown in Table \ref{tab:ic} 
have found to scale as $t_{\rm hll}:t_{\rm hllc}:t_{\rm hlld} =
1:1.4:1.8$.
This example demonstrates the effectiveness and strength of 
adopting a more complete Riemann solver when describing
the rich and complex features arising in relativistic
magnetized flows.

%

  
\subsection{Multidimensional Tests}
\label{sec:test23d}
%
%
%

We have implemented our $5$ wave Riemann solver into the framework
provided by the PLUTO code \citep{PLUTO}.
The constrained transport method is used to evolve the magnetic field.
We use the third-order, total variation diminishing Runge Kutta scheme 
together with piecewise linear reconstruction. 

\subsubsection{The 3D Rotor Problem}
%
%

\begin{figure}\begin{center}
 \includegraphics[width=0.5\textwidth]{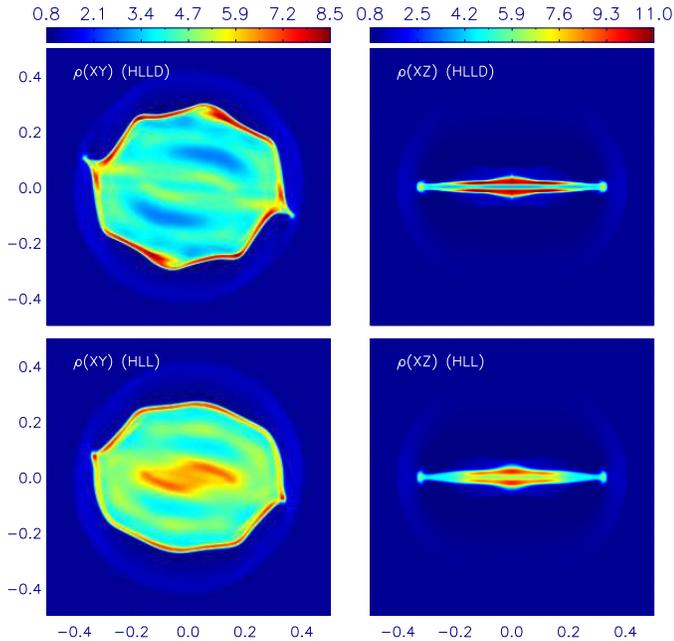}
 \caption{The 3D rotor test problem computed with HLLD 
          (top panels) and HLL (bottom panels) at
          the resolution of $256^3$. 
          Panels on the left show the density map (at $t=0.4$) in the 
          $xy$ plane at $z=0$ while panels to the right 
          show the density in the $xz$ plane at 
          $y = 0$.}
 \label{fig:rotor}
\end{center}\end{figure}
\begin{figure}\begin{center}
 \includegraphics[width=0.5\textwidth]{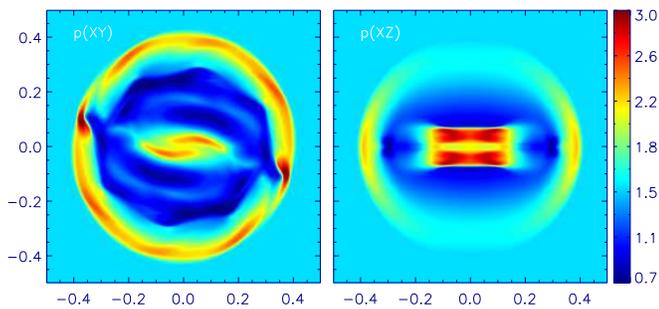}
 \caption{Same as Fig. \ref{fig:rotor} but showing the total 
          pressure in the $xy$ (left) and $xz$ (right) panels 
          for the HLLD solver.}
 \label{fig:rotor2}
\end{center}\end{figure}
\begin{figure}\begin{center}
 \includegraphics[width=0.5\textwidth]{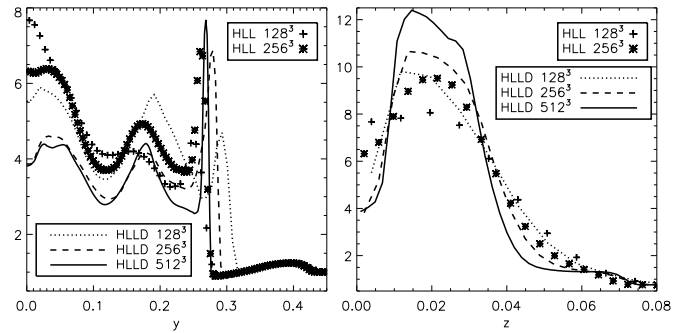}
 \caption{One dimensional cuts along the $y$ (left) and $z$ (right) 
          axis showing the density profiles at different resolutions
          ($128^3, 256^3$ and $512^3$) and with different solvers.
          Solid, dashed and dotted lines are used for the HLLD solver
          whereas plus and star symbols for HLL.}
 \label{fig:rotor3}
\end{center}\end{figure}

We consider a three dimensional version of the standard rotor problem
\citep{dZBL03}.
The initial condition consists of a sphere with radius 
$r_0 = 0.1$ centered at the origin of the domain taken to be
the unit cube $[-1/2,1/2]^3$.
The sphere is heavier ($\rho=10$) than the surrounding ($\rho=1$) and
rapidly spins around the $z$ axis with velocity components given by 
$(v^x, v^y, v^z) = \omega \left(-y,x,0\right)$ where $\omega = 9.95$ 
is the angular frequency of rotation.
Pressure and magnetic field are constant everywhere,
$p_g=1$, $\vec{B} = (1,0,0)$.

Exploiting the point symmetry, we carried computations until $t=0.4$ 
at resolutions of $128^3, 256^3$ and $512^3$ using both the HLLD and HLL solvers.
We point out that the HLLC of MB failed to pass this test, most likely
because of the flux-singularity arising in 3D computations in the zero 
normal field limit.

As the sphere starts rotating, torsional Alfv{\'e}n waves propagate outward 
carrying angular momentum to the surrounding medium.
The spherical structure gets squeezed into a disk configuration in 
the equatorial plane ($z=0$) where the two collapsing poles collide 
generating reflected shocks propagating vertically in the upper  
and lower half-planes.
This is shown in the four panels in Fig. \ref{fig:rotor} showing the density 
map in the $xy$ and $xz$ planes obtained with HLLD and HLL and in Fig. 
\ref{fig:rotor2} showing the total pressure.
After the impact a hollow disk enclosed by a higher density shell 
at $z=\pm 0.02$ forms (top right panels in Fig \ref{fig:rotor}).
In the $xy$ plane, matter is pushed in a thin, octagonal-like shell 
enclosed by a tangential discontinuity and what seems to be a 
slow rarefaction.
The whole configuration is embedded in a spherical fast rarefaction 
front expanding almost radially.
Flow distortions triggered by the discretization on a Cartesian 
grid are more pronounced with HLLD since we expect it to be
more effective in the growth of small wavelength modes.

In Fig. \ref{fig:rotor3} we compare the density profiles 
on the $y$ and $z$ axis for different resolutions and schemes.
From both profiles, one can see that the central region tends to 
become more depleted as the resolution increases.
Inspecting the profiles in the $y$ direction (left panel), we observe
that HLL and HLLD tend to under- and over-estimate (respectively) the 
speed of the thin density shell when compared to the reference solution 
computed with the HLLD solver at a resolution of $512^3$.
The height of the shell peak is essentially the 
same for both solvers, regardless of the resolution.

On the contrary, the right panel of Fig. \ref{fig:rotor3} 
shows a similar comparison along the vertical $z$ axis.
At the same resolution, HLL under-estimates the density peak located at 
$z=0.02$ and almost twice the number of grid zones is needed to match 
the results obtained with the HLLD solver. 
The location of the front is approximately the same regardless of the solver.

In terms of computational cost, integration carried with the HLLD
solver took approximately $\sim 1.6$ that of HLL.
This has to be compared with the CPU time required by HLL to reach a comparable 
level of accuracy which, doubling the resolution, would result
in a computation $\sim 2^4$ as long.
In this respect, three dimensional problems like the one considered here
may prove specially helpful in establishing the trade off
between numerical efficiency and accuracy which, among other things,
demand choosing between accurate (but expensive) solvers versus more 
diffusive (cheap) schemes.



\subsubsection{Kelvin-Helmholtz Unstable Flows}
%
%

\begin{figure}
 \includegraphics[width=0.5\textwidth]{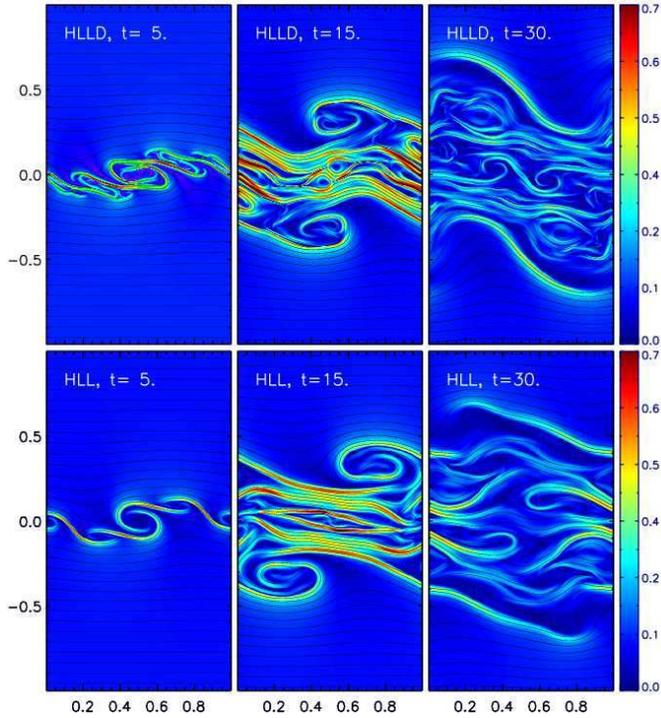}
 \caption{Color scale maps of $\sqrt{B_x^2 + B_y^2}/B_z$ 
          at different integration times, $t=5,15,30$.
          Panels on top (bottom) refer to computations accomplished 
          with HLLD (HLL). Poloidal magnetic field lines overlap.}
 \label{fig:kh}
\end{figure}

The setup, taken from \cite{BdZ06}, consists of a 2D planar
Cartesian domain, $x\in [0,1], y\in[-1,1]$ with a shear 
velocity profile given by
\begin{equation}
 v^x = -\frac{1}{4}\tanh\left(100\, y\right) \,.
\end{equation}
Density and pressure are set constant everywhere and initialized to 
$\rho = 1$, $p_g = 20$, while magnetic field components 
are given in terms of the poloidal and toroidal magnetization parameters 
$\sigma_{\rm pol}$ and $\sigma_{\rm tor}$ as
\begin{equation}
 \left(B^x, B^y, B^z\right) = \left(\sqrt{2\sigma_{\rm pol}p_g},
 0,  \sqrt{2\sigma_{\rm tor}p_g}\right) \,,
\end{equation}
where we use $\sigma_{\rm pol} = 0.01$, $\sigma_{\rm tor} = 1$.
The shear layer is perturbed by a nonzero component of the velocity,
\begin{equation}
 v^y = \frac{1}{400}\sin\left(2\pi x\right)\exp\left[-\frac{y^2}{\beta^2}\right]\,,
\end{equation}
with $\beta = 1/10$, while we set $v^z = 0$. 
Computations are carried at low (L, $90\times 180$ zones), medium 
(M, $180\times 360$ zones) and high (H,  $360\times 720$ zones) 
resolution.

\begin{figure}
 \includegraphics[width=0.5\textwidth]{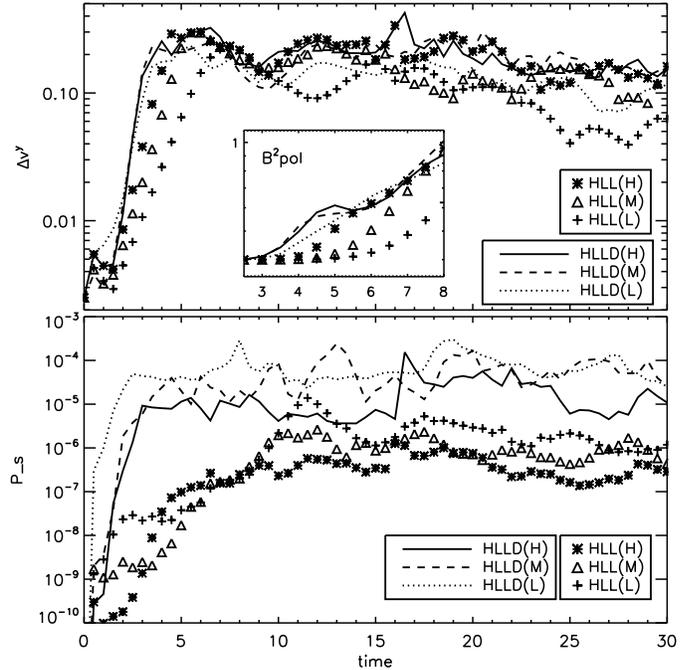}
 \caption{Top: growth rate (as function of time) 
          for the Kelvin-Helmholtz test problem
          computed as $\Delta v^y\equiv (v^y_{\max}-v^y_{\min})/2$
          at low (L), medium (M) and high (H) resolutions.
          Solid, dashed and dotted lines show results pertaining 
          to HLLD, whereas symbols to HLL.
          Bottom: small scale power as a function of time 
          for the Kelvin-Helmholtz application test.
          Integrated power is given by 
          $P_s = 1/2\int_{k_s/2}^{k_s}\int_{-1}^{1} |V(k,y)|^2 dy\, dk$ where
          $V(k,y)$ is the complex, discrete Fourier transform of 
          $v^y(x,y)$
          taken along the $x$ direction. Here $k_s$ is the
          Nyquist critical frequency.}
 \label{fig:kh2}
\end{figure}

For $t \lesssim 5$ the perturbation follows a linear growth phase 
leading to the formation of a multiple vortex structure. 
In the high resolution (H) case, shown in Fig \ref{fig:kh}, we observe
the formation of a central vortex and two neighbor, more stretched ones.
These elongated vortices are not seen in the computation of 
\cite{BdZ06} who employed the HLL solver at our medium resolution.
As expected, small scale patterns are best spotted with the HLLD solver, 
while tend to be more diffused using the two-wave HLL scheme.
The growth rate (computed as $\Delta v^y\equiv (v^y_{\max}-v^y_{\min})/2$, see
top panel in Fig. \ref{fig:kh2}), is closely related to the poloidal field
amplification which in turn proceeds faster for smaller numerical 
resistivity (see the small sub-plot in the same panel) and thus for
finer grids.
Still, computations carried with the HLLD solver at low (L),
medium (M) and high (H) resolutions reveal surprisingly similar 
growth rates and reach the saturation phase at essentially the same
time ($t\approx 3.5$). On the contrary, the saturation phase and the growth 
rate during the linear phase change with resolution when the HLL scheme is 
employed.

Field amplification is prevented by reconnection events during which 
the field wounds up and becomes twisted by turbulent dynamics.
Throughout the saturation phase (mid and right panel in Fig \ref{fig:kh}) 
the mixing layer enlarges and the field lines thicken into filamentary 
structures.
Small scale structure can be quantified by considering the power residing at
large wave numbers in the discrete Fourier transform of any flow quantity
(we consider the $y$ component of velocity).
This is shown in the bottom panel of Fig \ref{fig:kh2} where 
we plot the integrated power between $k_s/2$ and $k_s$ as function of time
($k_s$ is the Nyquist critical frequency).
Indeed, during the statistically steady flow regime ($t\gtrsim 20$),
the two solvers exhibits small scale power that differ by more than one order 
of magnitude, with HLLD being in excess of $10^{-5}$ (at all resolutions) 
whereas HLL below $10^{-6}$. 

In terms of CPU time, computations carried out with HLLD (at medium
resolution) were $\sim 1.9$ slower than HLL.

\subsubsection{Axisymmetric Jet Propagation}
%
%

\begin{figure}
 \includegraphics[width=0.5\textwidth]{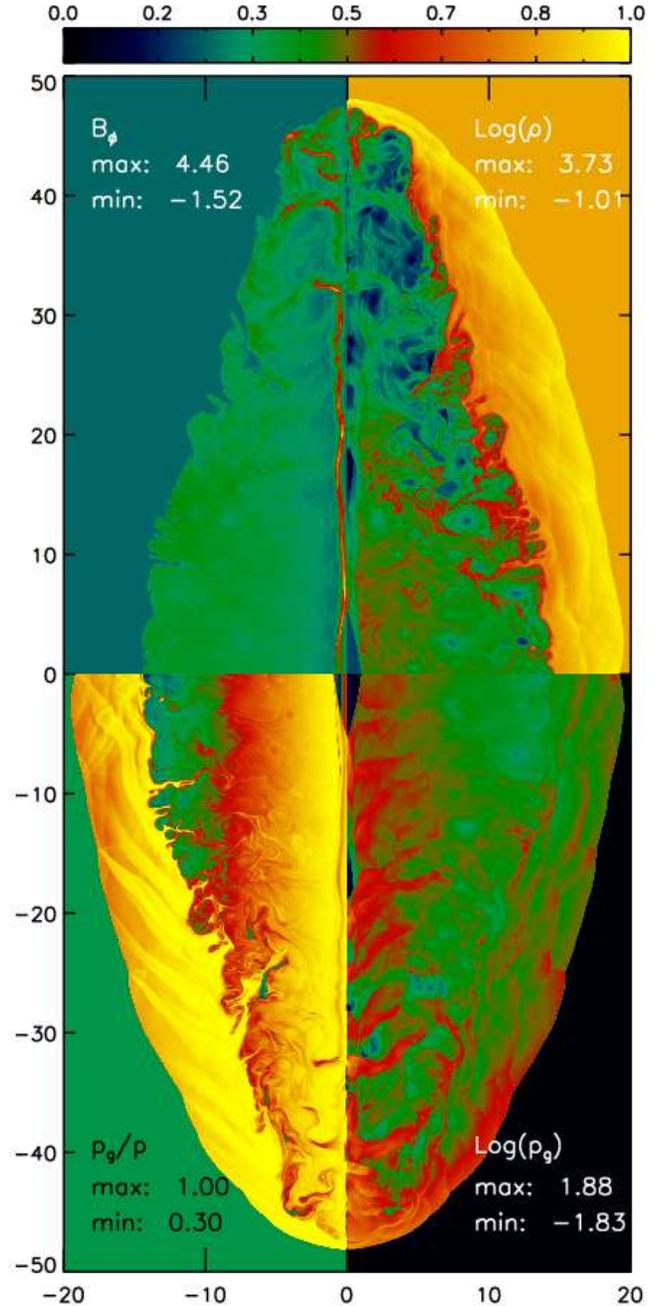}
 \caption{Left: composite color map image of the jet at $t=270$
          at the resolution of $40$ points per beam radius. 
          In clockwise direction, starting from the top right quadrant:
          density logarithm, gas pressure logarithm, thermal to total
          pressure ratio and $\phi$ component of magnetic field.
          The color scale has been normalized such that the maximum 
          and minimum values reported in each subplots correspond 
          to $1$ and $0$.}
 \label{fig:jet}
\end{figure}
\begin{figure}
 \includegraphics[width=0.5\textwidth]{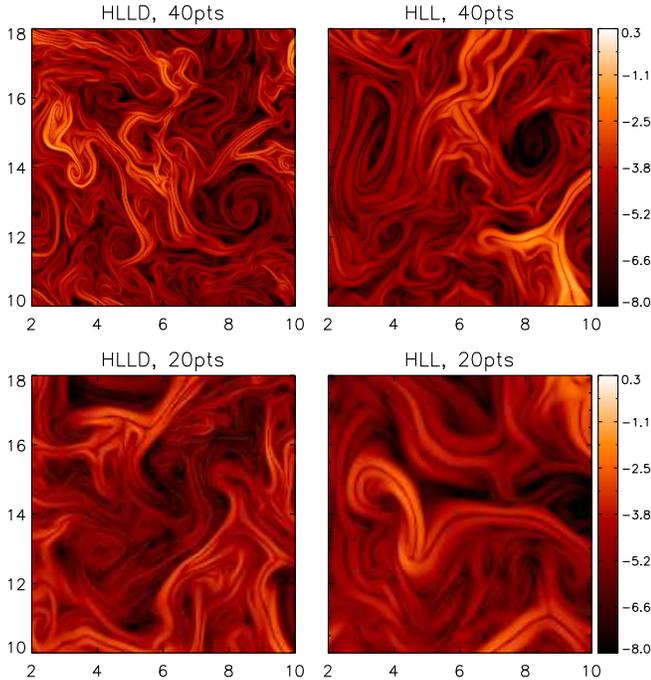}
 \caption{Enlargement of the turbulent flow region 
          $[2,10]\times[10,18]$ at
          $t=300$ showing the poloidal magnetic field structure
          (in log scale) for the high and medium resolution runs
          ($40$ and $20$ points per beam radius.}
 \label{fig:jet2}
\end{figure}
\begin{figure}
 \includegraphics[width=0.5\textwidth]{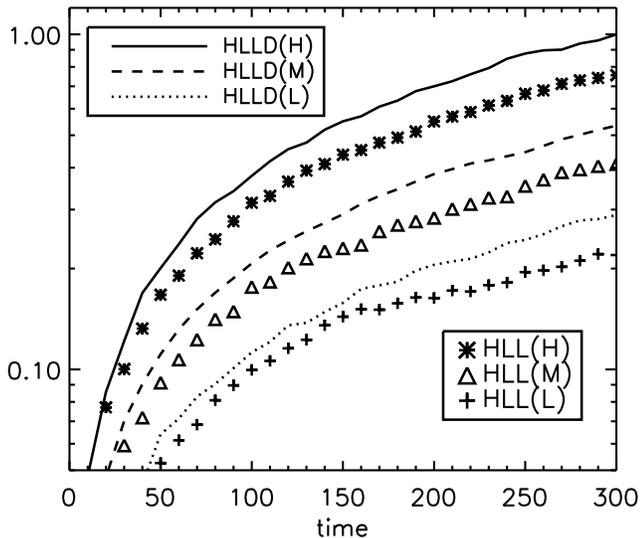}
 \caption{Volume average of $\nabla \vec{B}^2_p/\vec{B}^2_p$
          as a function of time. Here 
          $\vec{B}_p$ is the poloidal magnetic field.
          Solid, dashed and dotted lines 
          refers to computations carried out with HLLD, whereas symbols
          give the corresponding results obtained with HLL.}
 \label{fig:jet3}
\end{figure}

As a final example, we consider the propagation of a relativistic 
magnetized jet.
For illustrative purposes, we restrict our attention 
to axisymmetric coordinates with $r\in[0,20]$ and $z\in [0,50]$. 
The jet initially fills the region $r,z \le 1$ with density $\rho_j = 1$ and
longitudinal ($z$) velocity specified by $\gamma_j = 10$ 
($v^r = v^\phi = 0$).

The magnetic field topology is described by a constant 
poloidal term, $B^z$, threading both the jet and the ambient
medium and by a toroidal component $B^\phi(r) = \gamma_j b_\phi(r)$ with 
\begin{equation}
 b_\phi(r) = \left\{\begin{array}{l}
 b_m r/a \qquad\textrm{for}\qquad r < a      \,, \\ \noalign{\medskip}
 b_m a/r \qquad\textrm{for}\qquad a < r < 1  \,,
\end{array}\right.
\end{equation}
where $a=0.5$ is the magnetization radius and $b_m$ is a constant
and vanishes outside the nozzle.
The thermal pressure distribution inside the jet is set by the
radial momentum balance, $r\partial_r p_g = - b_\phi\partial_r(r b_\phi)$
yielding
\begin{equation}
 p_g(r) = p_j + b_m^2\left[1 - \min\left(\frac{r^2}{a^2},1\right)\right]\,,
\end{equation}
where $p_j$ is the jet/ambient pressure at $r=1$ and is 
recovered from the definition of 
the Mach number, $M=v_j\sqrt{\rho_j/(\Gamma p_j) + 
1/(\Gamma-1)}$, with $M=6$ and $\Gamma = 5/3$, 
although we evolve the equations using the 
approximated Synge gas equation of state of \cite{MmK07}.

The relative contribution of the two components is quantified 
by the two average magnetization parameters
$\sigma_z  \equiv B_z^2/(2\left<p_g\right>)$
$\sigma_\phi \equiv \left<{b^2_\phi}\right>/(2\left<p_g\right>)$
yielding
\begin{equation}
b_m = \sqrt{\frac{-4p_j\sigma_\phi}{a^2(2\sigma_\phi - 1 + 4\log a)}}
,\,
 B_z = \sqrt{\sigma_z\left(b_m^2a^2 + 2p_j\right)} \,,
\end{equation}
where for any quantity $q(r)$, $\left<{q}\right>$ gives 
the average over the jet beam $r\in[0,1]$.
We choose $\sigma_\phi = 0.3$, $\sigma_z = 0.7$, thus corresponding
to a jet close to equipartition.

The external environment is initially static ($\vec{v}_e=0$), heavier
with density $\rho_e=10^3$ and threaded only 
by the constant longitudinal field $B^z$. Pressure
is set everywhere to the constant value $p_j$.

We carry out computations at the resolutions of $10,20$ and $40$ zones 
per beam radius ($r=1$) and follow the evolution until $t=300$. 
The snapshot in Fig. \ref{fig:jet} shows the solution computed 
at $t=300$ at the highest resolution.

The morphological structure is appreciably affected by the magnetic 
field topology and by the ratio of the magnetic energy density to the 
rest mass, $b^2_\phi/\rho  \approx 0.026$.
The presence of a moderately larger poloidal component and a 
small Poynting flux favor the formation of a hammer-like structure 
rather than a nose cone \cite[see][]{Leis05, MMB05}.
At the termination point, located at $z\approx 40.5$, the beam strongly
decelerates and expands radially promoting vortex emission
at the head of the jet.

Close to the axis, the flow remains well collimated and undergoes a series 
of deceleration/acceleration events through a series of conical shocks,
visible at $z\approx 4.5, 19, 24, 28, 32$.
Behind these recollimation shocks, the beam strongly decelerates
and magnetic tension promotes sideways deflection of shocked material 
into the cocoon.

The ratio $p_g/p$ (bottom left quadrant in Fig \ref{fig:jet})
clearly marks the Kelvin-Helmholtz unstable
slip surface separating the backflowing, magnetized beam material 
from the high temperature (thermally dominated) shocked ambient medium.
In the magnetically dominated region turbulence dissipate magnetic
energy down to smaller scales and mixing occurs.
 The structure of the contact discontinuity observed in the
figures does not show suppression of KH instability. 
This is likely due to the larger growth of the toroidal field
component over the poloidal one \citep{Keppens08}.
However we also think that the small density ratio ($10^{-3}$) may favor
the growth of instability and momentum transfer through entrainment
of the external medium \citep{Rossi08}.

For the sake of comparison, we also plot (Fig \ref{fig:jet2}) 
the magnitude of the poloidal magnetic field in the region 
$r\in [2,10]$, $z\in [10,18]$ where turbulent patterns have
developed.
At the resolution of $40$ points per beam radius, HLLD discloses
the finest level of small scale structure, 
whereas HLL needs approximately twice the resolution 
to produce similar patterns.
This behaviour is quantitatively expressed,
in Fig \ref{fig:jet3}, by averaging the gradient $\log(B_r^2 + B_z^2)$
over the volume. Roughly speaking, HLL requires a resolution
$\sim 1.5$ that of HLLD to produce pattern with similar 
results.

\section{Conclusions}
\label{sec:conclusions}
%
%
%

A five-wave HLLD Riemann solver for the equations of 
relativistic magnetohydrodynamics has been presented.
The solver approximates the structure of the Riemann fan 
by including fast shocks, rotational modes and the contact 
discontinuity in the solution.
The gain in accuracy comes at the computational cost of
solving a nonlinear scalar equation in the total pressure.
As such, it better approximates Alfv\`en waves and we also
found it to better capture slow shocks and compound waves.
The performance of the new solver has been tested against
selected one dimensional problems, showing better accuracy 
and convergence properties than previously known 
schemes such as HLL or HLLC.

Applications to multi-dimensional problems have been presented
as well. The selected tests disclose better resolution of small
scale structures together with reduced dependency on grid resolution.
We argue that three dimensional computations may actually benefit 
from the application of the proposed solver which, albeit more 
computationally intensive than HLL, still allows to recover 
comparable accuracy and resolution with a reduced number of grid zones.
Indeed, since a relative change $\delta$ in the mesh spacing  
results in a factor $\delta^4$ in terms of CPU time, this
may largely favour a more sophisticated solver over 
an approximated one.
This issue, however, need to receive more attention in forthcoming studies.

\section*{Acknowledgments}

%
%

\appendix

\section{Propagation of Rotational Discontinuities}
\label{app:Alfven}
%
%
%

Left and right states across a rotational discontinuity 
can be found using the results outlined in \S\ref{sec:Alfven}.
More specifically, we construct a family of solutions
parameterized by the speed of the discontinuity $K^x$
and one component of the tangential field on the right of
the discontinuity. Our procedure can be shown to be be
equivalent to that of \cite{K97}.
Specifically, one starts by assigning $\rho, p_g, \vec{v}, \vec{B}^t$ on the
left side of the front ($\vec{B}^t\equiv(0,B^y,B^z)$) 
together with the speed of the front, $K^x$. 
Note that $B^x$ cannot be freely assigned but must be determined
consistently from Eq. (\ref{eq:KvB}). 
Expressing $K^k$ ($k\neq x$) in terms of $v^k, B^k$ and $B^x$
and substituting back in the $x-$ component of (\ref{eq:KvB}), 
one finds that there are two possible values of $B^x$ satisfying 
the quadratic equation
\begin{equation}
 a(B^x)^2 + bB^x + c = 0 \,,
\end{equation}
where the coefficients of the parabola are
\begin{equation}
 a = \eta - \frac{\left(\eta - K^xv^x\right)^2}{(K^x-v^x)^2}
\,,\quad
 b = 2\chi \left( v^x + 
     \frac{\eta - K^xv^x}{K^x - v^x}\right)
\,,\quad
\end{equation}
and
\begin{equation}
 c = w_g + \frac{\vec{B}^t\cdot\vec{B}^t}{\gamma^2}\,,
\end{equation}
with $\eta = 1 - (v^y)^2 - (v^z)^2$, $\chi = v^yB^y + v^zB^z$ and
$\gamma$ being the Lorentz factor.
The transverse components of $\vec{K}$ are computed as 
\begin{equation}
 K^{y,z} = v^{y,z} + \frac{B^{y,z}}{B^x}\left(K^x - v^x\right)\,.
\end{equation}

On the right side of the front, one has that $\rho$, $p_g$, $w$,
$B^x$ and $\vec{K}$ are the same, see \S\ref{sec:Alfven}.
Since the transverse field is elliptically polarized \citep{K97}, there are 
in principle infinite many solutions and one has the freedom to  
specify, for instance, one component of the field (say $B^y_R$).
The velocity $\vec{v}_R$ and the $z$ component of the field 
can be determined in the following way.
First, use Equation (\ref{eq:vc}) to express $v^k_{cL}$
($k = x,y,z$) as function of $B^z_R$ for given $B^x_R$ and $B^y_R$.
Using the jump condition for the density together 
with the fact that $\rho$ is invariant, we solve the 
nonlinear equation
\begin{equation}
 \rho_L\gamma_L\left(K^x - v^x_L\right) = 
 \rho_R\gamma_R\left(K^x - v^x_R\right) \,,
\end{equation}
whose roots gives the desired value of $B^z_R$.

\label{lastpage}

\begin{thebibliography}{}

\bibitem[\protect\citeauthoryear{Anile \& Pennisi}{1987}]{AP87}
  Anile, M., \& Pennisi, S.   \
  1987, Ann. Inst. Henri Poincar{\'{e}}, 46, 127

\bibitem[\protect\citeauthoryear{Anile}{1989}]{Anile89}
  Anile, A.~M.\
  1989, Relativistic Fluids and Magneto-fluids (Cambridge: Cambridge University Press), 55

\bibitem[\protect\citeauthoryear{Balsara}{2001}]{Balsara01}
  Balsara, D.~S.\ 
  2001, ApJS, 132, 83 

\bibitem[\protect\citeauthoryear{Berger \& Colella}{1989}]{BC89}
  Berger, M. J.  and Colella, P. \
  J. Comput. Phys., 82, pp. 64-84, 1989.


\bibitem[\protect\citeauthoryear{Brio \& Wu}{1988}]{BW88}
  Brio, M., \& WU, C.-C. \
  1988, J. Comput. Phys., 75, 400

\bibitem[\protect\citeauthoryear{Bucciantini \& Del Zanna}{2006}]{BdZ06}
  Bucciantini, N., \& Del Zanna, L.
  2006, Astronomy \& Astrophysics, 454, 393

\bibitem[\protect\citeauthoryear{Davis}{1988}]{Davis88}
  S.F. Davis, SIAM J. Sci. Statist. Comput. 9 (1988) 445.

\bibitem[\protect\citeauthoryear{Einfeldt et al.}{1991}]{EMRS91}
  Einfeldt, B., Munz, C.D., Roe, P.L., and Sj{\" o}green, B.\
  1991, J. Comput. Phys., 92, 273

\bibitem[\protect\citeauthoryear{Del Zanna et al.}{2003}]{dZBL03}
  Del Zanna, L., Bucciantini, N., \& Londrillo, P.\
  2003, Astronomy \& Astrophysics, 400, 397 (dZBL)

\bibitem[\protect\citeauthoryear{Del Zanna et al.}{2007}]{dZZB07} 
  Del Zanna, L., Zanotti, O., Bucciantini, N., \& Londrillo, P.\ 
  2007, Astronomy \& Astrophysics, 473, 11 

\bibitem[\protect\citeauthoryear{Gammie et al.}{2003}]{GKT03} 
  Gammie, C.~F., McKinney, J.~C., \& T{\'o}th, G.\ 
  2003, The Astrophysical Journal, 589, 444 

\bibitem[\protect\citeauthoryear{Gehmeyr et al.}{1997}]{GCM97}
  Gehmeyr, M., Cheng, B., \& Mihalas, D.\
  1997, Shock Waves, 7, 255

\bibitem[\protect\citeauthoryear{Giacomazzo \& Rezzolla}{2006}]{GR06}
  Giacomazzo, B., \& Rezzolla, L.\ 2006, 
  Journal of Fluid Mechanics, 562, 223 

\bibitem[\protect\citeauthoryear{Gurski}{2004}]{Gurski04}
  Gurski, K.F. \
  2004, SIAM J. Sci. Comput, 25, 2165

\bibitem[\protect\citeauthoryear{Harten et al.}{1983}]{HLL83}
  Harten, A., Lax, P.D., and van Leer, B.\
  1983, SIAM Review, 25(1):35,61

\bibitem[\protect\citeauthoryear{Honkkila \& Janhunen}{2007}]{HJ07}
  Honkkila, V., \& Janhunen, P.\ 2007, 
  Journal of Computational Physics, 223, 643 

\bibitem[\protect\citeauthoryear{Jeffrey \& Taniuti}{1964}]{JT64}
  Jeffrey A., Taniuti T., \
  1964, Non-linear wave propagation. Academic Press, New York

\bibitem[\protect\citeauthoryear{Keppens et al.}{2008}]{Keppens08}
  Keppens, R., Meliani, Z., van der Holst, B., \& Casse, F.\ 
  2008, Astronomy \& Astrophysics, 486, 663 


\bibitem[\protect\citeauthoryear{Koldoba et al.}{2002}]{KKU02}
   Koldoba, A.~V., Kuznetsov, O.~A., \& Ustyugova, G.~V.\ 
   2002, MNRAS, 333, 932 

\bibitem[\protect\citeauthoryear{Komissarov}{1997}]{K97}
  Komissarov, S.~S.\
  1997, Phys. Lett. A, 232, 435

\bibitem[\protect\citeauthoryear{Komissarov}{1999}]{K99}
  Komissarov, S.~S.\
  1999, mnras, 308, 1069

\bibitem[\protect\citeauthoryear{Leismann et al.}{2005}]{Leis05} 
  Leismann, T., Ant{\'o}n, L., Aloy, M.~A., M{\"u}ller, E., Mart{\'{\i}}, J.~M., Miralles, J.~A., \& Ib{\'a}{\~n}ez, J.~M.\ 
  2005, Astronomy \& Astrophysics, 436, 503 

\bibitem[\protect\citeauthoryear{Li}{2005}]{Li05}
  Li S., 
  2005, J. Comput. Phys., 344-357

\bibitem[\protect\citeauthoryear{Lichnerowicz}{1976}]{L76}
  Lichnerowicz, A.\
  1976, Journal of Mathematical Physics, 17, 2135

\bibitem[\protect\citeauthoryear{Lichnerowicz}{1967}]{lich67}
  Lichnerowicz, A.\ 1967, Relativistic Hydrodynamics and Magnetohydrodynamics,
  New York: Benjamin, 1967

\bibitem[\protect\citeauthoryear{Mignone \& Bodo}{2005}]{MB05}
  Mignone, A., \& Bodo, G.\ 2005, MNRAS, 364, 126 

\bibitem[\protect\citeauthoryear{Mignone et al.}{2005}]{MMB05} 
  Mignone, A., Massaglia, S., \& Bodo, G.\ 2005, Space Science Reviews, 121, 21 

\bibitem[\protect\citeauthoryear{Mignone \& Bodo}{2006}]{MB06}
  Mignone, A., \& Bodo, G.\ 2006, MNRAS, 368, 1040 (MB)

\bibitem[\protect\citeauthoryear{Mignone et al.}{2007}]{PLUTO}
  Mignone, A., Bodo, G., Massaglia, S., Matsakos, T., Tesileanu, O., Zanni, C., 
  \& Ferrari, A.\ 
  2007, Astrophysical Journal Supplement, 170, 228 

\bibitem[\protect\citeauthoryear{Mignone \& McKinney}{2007}]{MmK07}
  Mignone, A., \& McKinney, J.~C.\ 
  2007, MNRAS, 378, 1118 

\bibitem[\protect\citeauthoryear{Miyoshi \& Kusano}{2005}]{MK05} 
  T. Miyoshi, K. Kusano, K.,
  J. Comp. Phys. 208 (2005) 315 (MK)

\bibitem[\protect\citeauthoryear{Noh}{1987}]{Noh87}
  Noh, W.F.\
  1987, J. Comput. Phys., 72,78

\bibitem[\protect\citeauthoryear{Press et al.}{1992}]{NumRec}
  Press, W., S. Teukolsky, W. Vetterling, and B. Flannery (1992). 
  Numerical Recipes in C (2nd ed.). Cambridge, UK: Cambridge University Press.

\bibitem[\protect\citeauthoryear{Romero et al.}{2005}]{Rom05} 
  Romero, R., Mart{\'{\i}}, J.~M., Pons, J.~A., Ib{\'a}{\~n}ez, J.~M., \& Miralles, J.~A.\ 
  2005, Journal of Fluid Mechanics, 544, 323 

\bibitem[\protect\citeauthoryear{Rossi et al.}{2008}]{Rossi08}
  Rossi, P., Mignone, A., Bodo, G., Massaglia, S., \& Ferrari, A.\ 
  2008, Astronomy \& Astrophysics, 488, 795 

\bibitem[\protect\citeauthoryear{Toro et al.}{1994}]{TSS94}
  Toro, E.~F., Spruce, M., and Speares, W.\ 
  1994, Shock Waves, 4, 25

\bibitem[\protect\citeauthoryear{Toro}{1997}]{Toro97}
  Toro, E.~F.\
  1997, Riemann Solvers and Numerical Methods for Fluid Dynamics,
  Springer-Verlag, Berlin

\bibitem[\protect\citeauthoryear{van der Holst et al.}{2008}]{vdHKM08} 
  van der Holst, B., Keppens, R., \& Meliani, Z.\ 
  2008, arXiv:0807.0713 



\end{thebibliography}
\end{document}